\documentclass[reprint, amsmath, amssymb, aps, pra, floatfix]
{revtex4-2}
\usepackage{graphicx}
\usepackage{braket}
\usepackage{blkarray}
\usepackage{upgreek}
\usepackage{color}
\usepackage[normalem]{ulem}
\usepackage[colorlinks=true, urlcolor=blue, linkcolor=red]{hyperref}

\newcommand{\tjvw}{$t\mbox{-}J\mbox{-}V\mbox{-}W$ }

\begin{document}

\title{Ultracold polar molecules for quantum science and technology}
\author{Philip D. Gregory}
\author{Simon L. Cornish}
\affiliation{Department of Physics, Durham University, South Road, Durham, DH1 3LE, UK.}

\begin{abstract}
Ultracold polar molecules present a rich platform for experimental quantum science leveraging their unique properties. Of particular relevance is the intrinsic electric dipole moment of the molecule, which results in strong coupling to microwave fields and can be used to engineer long-range, anisotropic, and tunable dipole-dipole interactions. Such interactions can generate entanglement between molecules and enable the study of novel quantum many-body phenomena in parameter regimes inaccessible in other systems. Modern experiments in optical tweezers and optical lattices have demonstrated controlled interactions between molecules in coherent quantum states, and recently bulk gases of molecules have been cooled to quantum degeneracy. Motivated by these landmark advances, we present this article to explain many of the key concepts relevant to current research in the field. 
\end{abstract}

\maketitle

\noindent

The cooling of gases of neutral atoms to temperatures within a millionth of a degree of absolute zero has transformed the study of quantum physics and had profound impact on society. Such ultracold temperatures enable exquisite control and, in turn, the exploration of many novel quantum phenomena. Since the early observations of Bose-Einstein condensation over thirty years ago~\cite{Anderson1995}, a constant theme in the field has been to push this control to more and more complex systems, with each advance opening up new opportunities for quantum science and technology. Examples include the extension to Fermionic atoms~\cite{Fermi_review}, the introduction of magnetically tunable Feshbach resonances~\cite{Chin2010}, confinement in optical lattices~\cite{Lattices_review} and reduced dimensions~\cite{Lattices_review}, experiments with multiple species~\cite{Multispecies_review}, the cooling of highly magnetic atoms~\cite{Magneticatoms_review}, engineering artificial gauge fields~\cite{Gaugefield_review} and excitation to Rydberg states~\cite{Rydberg1_review,Rydberg2_review}. In the same vein, there has been a strong and sustained effort to produce \emph{ultracold~molecules}.

Ultracold polar molecules uniquely combine a rich internal structure of long-lived hyperfine, rotational and vibrational states with access to long-range, anisotropic, and tunable dipole-dipole interactions. These properties have stimulated many proposed applications spanning ultracold chemistry~\cite{Karman2024}, precision measurement~\cite{DeMille2024}, quantum simulation and quantum computation~\cite{Cornish2024}. Following remarkable experimental progress over the last decade~\cite{Gregory2026}, many of these applications are now within reach. Therefore, this article aims to provide a timely and accessible introduction to the key concepts that underpin modern experiments with ultracold molecules. We focus on the physics of relatively simple heteronuclear diatomic molecules and explain how they can be used to construct artificial quantum materials for the study quantum many-body physics - an application commonly referred to as quantum simulation~\cite{Georgescu2014,Cornish2024}.

\begin{figure}[t!]
\centering
\includegraphics[width=\columnwidth]{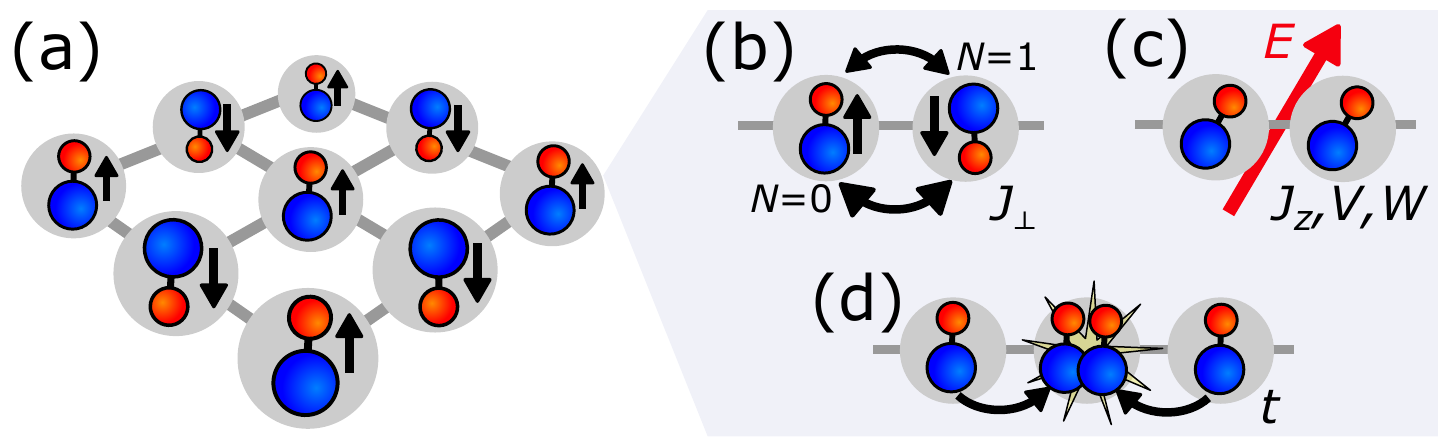}
\caption{Quantum simulation with arrays of ultracold polar molecules. (a)~An array of polar molecules acts as a many-body system of dipolar spins. Such an array can be prepared using the periodic potentials of optical lattices or using arrays of tightly-focused optical tweezers. Different spin orientations (i.e. up and down) can be mapped onto different rotational states of the molecules. Molecules in such arrays can interact via (b)~dipolar spin exchange, (c)~space-fixed dipole-dipole interactions, and (d)~collisions between pairs of molecules tunnelling onto the same potential minima.}
\label{fig:Introduction}
\end{figure}

The prototypical example of an ultracold molecule-based quantum simulator is illustrated in Fig.\,\ref{fig:Introduction}(a). The molecules are confined to periodic potential imposed by an optical lattice and initialized with a single molecule occupying each site. In the simplest case, each of the molecules is prepared in one of two internal states, usually different rotational states, that are treated as pseudo-spins. Crucially, molecules in different sites in the array can interact. These interactions arise from an intrinsic electric dipole moment in the molecule  that stems from the uneven sharing of electrons between the constituent atoms. As we will explain, pairs of molecules then experience a dipole-dipole interaction, which in the simplest case takes the form 
\begin{equation}
V_\mathrm{dd} = \frac{d^2}{4\pi\epsilon_0 r^3} (1-3\cos^2\Theta),
\label{eq:DipoleDipole}
\end{equation}
where $d$ is the magnitude of the dipole, $r$ is the distance between the molecules, $\Theta$ is the angle between the vector connecting the pair of molecules and the quantisation axis, and $\epsilon_0$ is the permittivity of free space. Such interactions are considered long range because they scale with the inter-molecular separation as $1/r^3$ and anisotropic because they vary with the angle $\Theta$. An important feature of ultracold polar molecules is that the dipole-dipole interactions can be accessed with the molecules in long-lived states of the electronic ground state, where spontaneous emission is negligible. As a consequence, long-range interactions can modify the motion of itinerant molecules leading to a range of novel quantum many-body phases.

Returning to the example in Fig.\,\ref{fig:Introduction}(a), we note that the interactions between molecules can arise in a variety of different ways. Molecules in different states connected by a transition dipole moment can exchange rotational energy via resonant dipolar spin exchange [Fig.\,\ref{fig:Introduction}(b)]. Alternatively, in an applied electric field, each molecule develops a static lab-frame dipole moment that influences the potential energy of its neighbours, with the sign and strength of the interactions determined by the direction and magnitude of the applied field [Fig.\,\ref{fig:Introduction}(c)]. Finally, the molecules can tunnel between the sites of the lattice to physically rearrange, and if entering the same site as one-another experience an on-site interaction or collision [Fig.\,\ref{fig:Introduction}(d)]. As we will show, this conceptually simple system allows access to remarkably rich quantum physics.

The article is structured as follows. In section I, we briefly outline the leading methods used to produce ultracold polar molecules. In section II, we  discuss in detail the rotational and hyperfine structure of diatomic molecules and explain how dipole-dipole interactions between molecules may be engineered. In section III, we study molecules in optical traps along with recent developments to engineer rotationally-magic traps that maximize quantum coherence between rotational states of the molecules. In section IV, we  examine collision loss in ultracold molecular gases and how losses may be suppressed using dipolar shielding. Finally, in section V we will conclude by explaining how, by combining all of the concepts introduced in the earlier sections, novel models of quantum many-body physics may be implemented with ultracold polar molecules.

\section{Production of ultracold diatomic molecules}

There are two leading approaches to the production of ultracold gases of polar molecules, which we refer to as \emph{direct} and \emph{indirect}. In the \emph{direct} approach a relatively hot gas of molecules is first produced and then directly laser cooled. In the \emph{indirect} approach, a pre-cooled mixture of atoms is first prepared. Atoms are then associated in this mixture to form molecules that inherit the ultracold temperature from their parent atomic species. Each technique has its own advantages and disadvantages, and each can only be applied to certain molecular species. Where possible, we will keep this review general, but many of the specific examples given later will necessarily focus on bialkali molecules that are produced by association. 

\subsection*{Direct laser cooling of molecules}

Laser cooling requires the scattering of many photons. The species of molecule must therefore be carefully selected such that there exists an accessible cycling transition that is sufficiently closed. That is, the molecule must be able to scatter a photon and return to the state in which it started the majority of the time. This is generally difficult because there are no selection rules that govern the transitions between different vibrational states of molecules; instead the transition probability is governed by the square of the overlap integral between the vibrational wave functions of the initial and final states -- known as the Franck-Condon Factor.  This means that molecules that are electronically excited can usually decay to multiple different vibrational levels of their electronic ground state. 

\begin{figure}[t!]
\centering
\includegraphics[width=0.6\columnwidth]{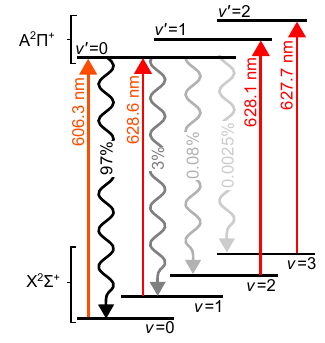}
\caption{Energy levels and transitions used for magneto-optical trapping of CaF inspired by Fig.\,1(a) of Williams~{\it et~al.}~\cite{Williams2017}. Straight arrows indicate driven transitions. Cooling and trapping in the MOT primarily takes place on the 606.3\,nm transition between the vibrational states $\mathrm{X}^2\Sigma^+ (v=0)$ and $\mathrm{A}^2\Pi^+(v'=0)$. Wavy arrows show pathways for spontaneous emission from $v'=0$, with the most likely pathway being back down to $v=0$ where the 606.3\,nm light can once again excite the molecule. Molecules that take a different path are repumped back down to $v=0$ by driving the 627.7\,nm, 628.1\,nm, 628.6\,nm transitions. }
\label{fig:CaF_MOT}
\end{figure}

Sufficiently-closed transitions have been found in a number of molecules, notably SrF~\cite{Shuman2010} and CaF~\cite{Truppe2017}. These molecules are characterised by highly diagonal, near-unity Franck-Condon Factors stemming from the fact that the optically active electron does not participate in the bond. Consequently, transitions can be driven with very little leakage into other states. This enables laser cooling and trapping with a small number of repump lasers to plug the leakage of population to undesired `dark' states; the laser scheme used for magneto-optical trapping of CaF is shown in Fig.~\ref{fig:CaF_MOT}. Note that rotational branching is avoided through angular momentum selection rules~\cite{BrownandCarrington}.

Beams of SrF and CaF molecules can be generated using buffer-gas sources. Here, SrF and CaF molecules are produced by laser ablation of strontium (Sr) or calcium (Ca) respectively in the presence of sulfur hexafluoride (SF$_6$) gas. These molecules are then pre-cooled to cryogenic temperatures via collisions with a helium buffer gas before leaving the source via a small aperture. The molecules produced by these sources can be laser slowed and then directly loaded into a magneto-optical trap as the starting point for experiments at ultracold temperatures~\cite{Barry2014,Norrgard2016,Truppe2017}. A much more detailed introduction to laser cooling of molecules can be found in~\cite{Tarbutt2018}.

\subsection*{Indirect approach to ultracold molecules}

\begin{figure}[t!]
\centering
\includegraphics[width=\columnwidth]{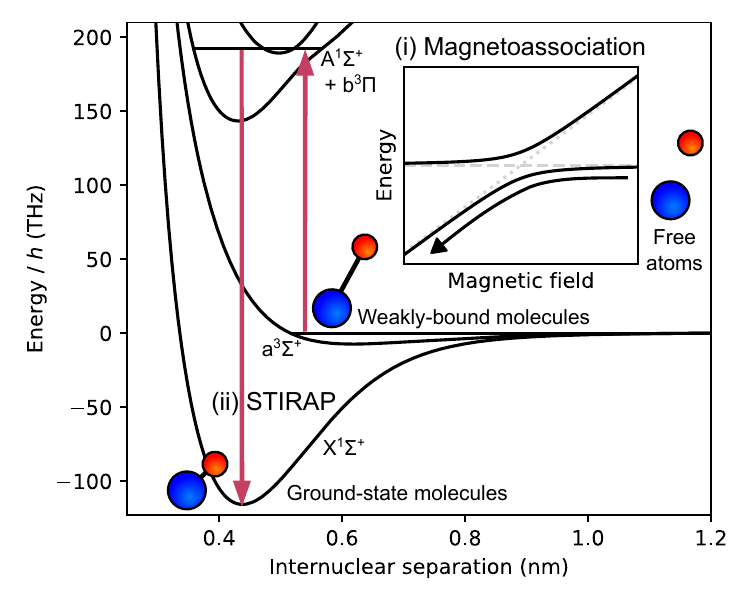}
\caption{Association scheme for converting ultracold atoms into polar molecules. The main plot shows the electronic potentials for the $^{87}$Rb$^{133}$Cs molecule as an example. The inset shows a simple 2-level energy-level picture of the first step (i)~magnetoassociation. Uncoupled states associated with the free atoms and a weakly-bound molecular state are shown by the dashed and dotted grey lines respectively, and the coupled states are shown by the black solid lines. The relative energies between these atomic and molecular states can be varied with an applied magnetic field, and where they become degenerate a Feshbach resonance is present due to the avoided crossing. By sweeping the magnetic field down across this resonance, pairs of atoms can adiabatically follow the lower coupled state to smoothly transition between atoms and weakly-bound molecules. These molecules are then transferred to their absolute ground state in the second step (ii)~STIRAP. This is achieved by coupling both the initial weakly-bound state and the ground state to a common electronically excited state by driving two optical transitions, which are indicated by the vertical red arrows. }
\label{fig:Association}
\end{figure}

The indirect approach to ultracold molecules circumvents the need for the molecules to possess their own cycling transitions, instead utilising the many developments in atomic laser cooling made over the last few decades. The approach is illustrated in~Fig.\,\ref{fig:Association}. It starts with an ultracold mixture of `free' atoms, usually held in an optical dipole trap, with high phase-space density. Weakly-bound molecules are produced from this mixture by applying a time-varying magnetic field that sweeps across an interspecies Feshbach resonance~\cite{Chin2010}. Such resonances occurs where the energy of the free atoms matches that of a molecular bound state that sits in a higher-lying molecular potential. For pairs of alkali atoms, resonant coupling between these states originates from the difference between the singlet and triplet potential curves and the magnetic dipolar interaction between the electron spins. This leads to an avoided crossing between the free atomic state and the molecular bound state at the Feshbach resonance. By ramping the magnetic field across the resonance pairs of atoms adiabatically follow the avoided crossing to form weakly-bound diatomic molecules. These weakly-bound molecules are then transferred to their electronic, vibrational, and rotational ground state optically using stimulated Raman adiabatic passage (STIRAP)~\cite{Vitanov2017}. 

This approach has been highly successful in producing dense gases of ultracold molecules at relatively high phase space densities from various combinations of alkali atoms. The first polar molecule produced using this approach was KRb~\cite{Ni2008}. Over the past decade this approach has advanced significantly leading to the first quantum degenerate samples of Fermionic KRb~\cite{DeMarco2019} and NaK~\cite{Schindewolf2022}, and Bosonic NaCs~\cite{Bigagli2024} and NaRb~\cite{Shi2025} molecules.

\section{Molecular Structure and Interactions}

\subsection*{Rigid rotor molecules}
We will consider here only diatomic molecules in the lowest vibrational level of the ground electronic potential. For heteronuclear diatomic molecules there exists an electric dipole moment that points along the internuclear axis, towards the element in the pair that has the lowest electronegativity. Although the direction of the dipole is well defined in the molecule frame, this is generally not the case in the laboratory (lab) frame!

The orientation of molecules in the lab frame is intimately linked to its rotational angular momentum. To understand the quantised rotational structure of the diatomic molecule we consider it as a simple rigid rotor, that is, two nuclei with masses $m_1, m_2$ connected together by a rigid bond of length $R$. This simple molecule can rotate end over end, and the Hamiltonian that describes this motion is
\begin{equation}
H_\mathrm{rot} = B_v \boldsymbol{N}^2,
\end{equation}
where $\boldsymbol{N}$ is the vector associated with the angular momentum, and $B_v$ is the rotational constant. The rotational constant is defined by $B_v=\hbar^2/(2mR^2)$ where $\hbar$ is the reduced Planck constant, $m$ is the reduced mass $m_1m_2/(m_1+m_2)$ and $R$ is the equilibrium separation between the nuclei, that we are assuming does not vary with time by neglecting vibration. Note, that even though we neglect vibration, $B_v$ depends on both the vibrational and electronic state of the molecule as this can affect the equilibrium internuclear separation. Defining the quantum number associated with the rotational angular momentum $N$, this Hamiltonian results in a structure of energy levels
\begin{equation}
U_\mathrm{rot} = B_v N (N+1).
\end{equation}
The rotational states of the molecule therefore form an anharmonic ladder, where the states become spaced further apart in energy as the rotational angular momentum is increased. 

\begin{figure}[t!]
\centering
\includegraphics[width=1.05\columnwidth]{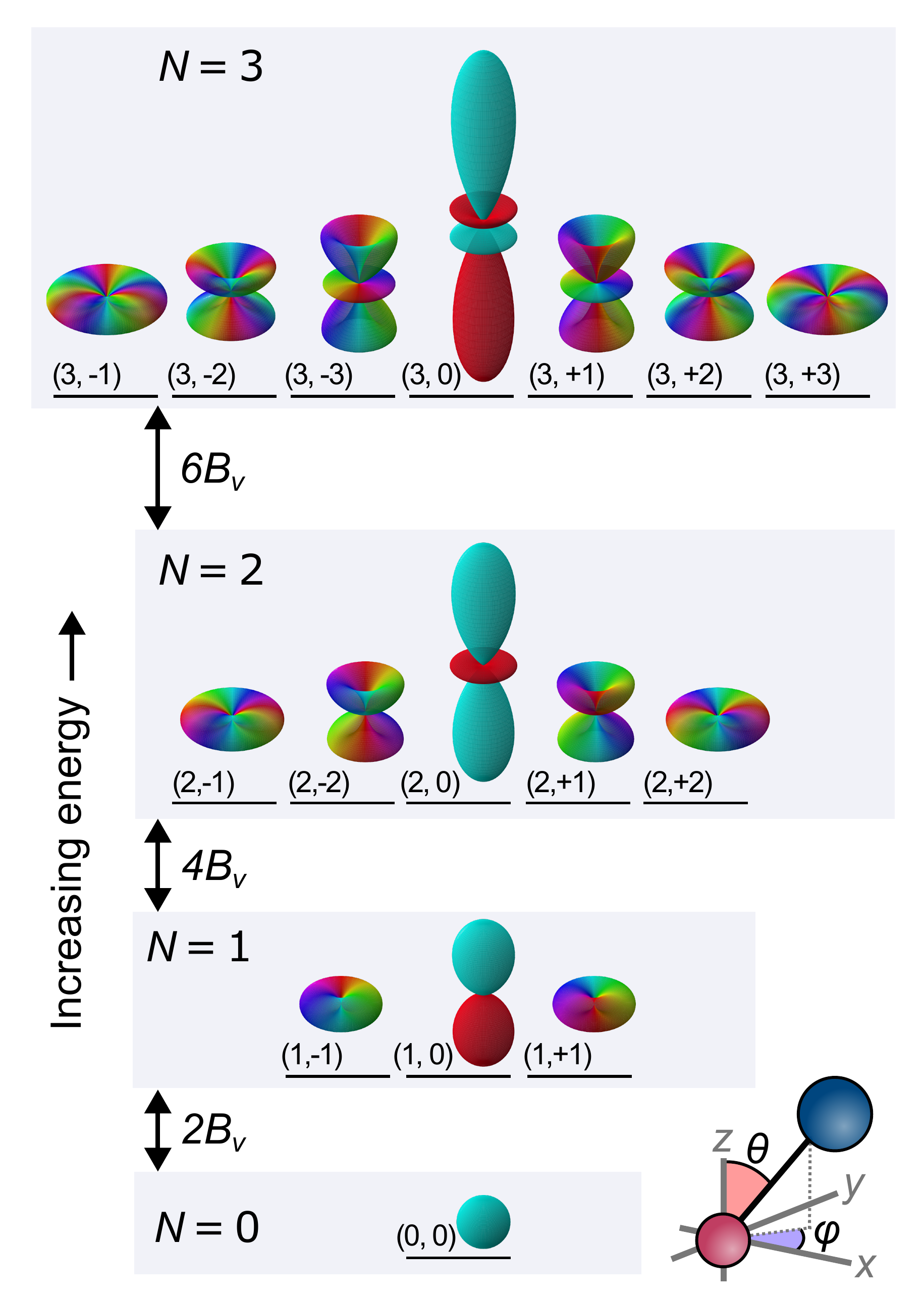}
\caption{The rotational states of the rigid-rotor molecule, labelled by $(N,M_N)$. The wavefunction association with each $(N,M_N)$ corresponds to the spherical harmonic~$Y^{M_N}_N(\theta,\phi)$. The probability distribution $\psi^*\psi$ is illustrated by the 3D plots, with the phase angle indicated by the surface colour. The angles $\theta, \phi$ correspond to the orientation of the intermolecular axis in the lab frame, and therefore the direction of the molecular dipole.}
\label{fig:RigidRotor}
\end{figure}

Solving the Schr\"{o}dinger equation for the rigid rotor yields wavefunctions that are described by spherical harmonics
\begin{equation}
\psi = Y_{N,M_N}(\theta, \phi),
\end{equation}
where $M_N$ is the projection of the rotational angular momentum $\boldsymbol{N}$ along the quantisation axis, and $\theta, \phi$ describe the orientation of the internuclear axis. We can now see that the orientation of the molecules is not well defined, but instead relies on the probability distribution $\psi^*\psi$. These distributions are shown for the rotational states up to $N=3$ in Fig.~\ref{fig:RigidRotor}. 

Importantly, for any wavefunction defined by a single $N, M_N$ component, the expectation value for the magnitude of the lab-frame electric dipole moment is zero. This is easiest to see for the rotational ground state $N=0$ where the probability distribution is spherical such that the molecule is equally likely to be oriented in any direction. As the lab-frame dipole moment is zero, molecules prepared in a single $(N, M_N)$ state do not experience any dipole-dipole interactions. Such interactions must be carefully engineered in experiments by applying either dc or ac (oscillating) electric fields to the molecules or by preparing the molecules in different rotational states that are connected by an electric-dipole-allowed transition.

\subsection*{Engineering lab-frame dipoles}

We will first examine the effect of applying a dc electric field to the molecules. The Hamiltonian that describes the dc Stark effect is 
\begin{equation}
H_\mathrm{dc}=-d \boldsymbol{E}\cdot \boldsymbol{\hat{n}},
\end{equation}
where $d$ is the molecule-frame dipole moment, $\boldsymbol{E}$ is the electric field vector, and $\boldsymbol{\hat{n}}$ is the unit vector pointing along the internuclear axis of the molecule. Fixing the electric field to point along $z$, which we define as the quantisation axis, this term becomes
\begin{equation}
H_\mathrm{dc}=-d E \cos(\theta),
\label{eq:dcHam}
\end{equation}
where $E$ is the magnitude of the electric field and $\theta$ is the angle of the internuclear axis with respect to $z$. There is now an energetic preference for the molecule to be oriented along the electric field, as this minimises $\theta$ and therefore minimises the potential energy associated with Eq.~\ref{eq:dcHam}. This system is analogous to a spherical pendulum, such that the eigenstates of this rigid rotor in an electric field are often referred to as `pendular' states. 

To calculate the effect of the electric field on the molecule, we calculate the matrix elements of Eq.~\ref{eq:dcHam} in the $(N, M_N)$ basis. As each of the wavefunctions is a spherical harmonic, the matrix elements are
\begin{multline}
\bra{N,M_M}H_\mathrm{dc}\ket{N',M'_N} = \\ 
-d E \int_{0}^{2\pi} \int_0^\pi Y^*_{N,M_N} \cos(\theta) Y_{N',M'_N}~\sin(\theta) d\theta d\phi. 
\end{multline}
Making the substitution $\cos(\theta)=Y_{1,0} \times \sqrt{4\pi/3}$ allows us to write the integral as the product of three spherical harmonics. Such products are well known and can be written as
\begin{multline}
\bra{N,M_N}H_\mathrm{dc}\ket{N',M'_N} = \\
-d E \sqrt{(2N+1)(2N'+1)} (-1)^{M_N} \\
\times \begin{pmatrix}
N & 1 & N'\\
-M_N & 0 & M'_N
\end{pmatrix}
\begin{pmatrix}
N & 1 & N'\\
0 & 0 & 0
\end{pmatrix},
\label{eq:dcMatrixElements}
\end{multline}
where the terms in brackets are Wigner-3j coefficients~\footnote{These coefficients can be easily evaluated using open source tools such as those found in the sympy.physics.wigner library for Python.}. These coefficients effectively define selection rules for which states become coupled; we find that the dc Stark effect mixes states that differ in $N$ by 1, but only mixes those states that have the same value of $M_N$. In matrix form, these are seen as off-diagonal elements that are non-zero. We illustrate this for the total Hamiltonian $H=H_\mathrm{rot}+H_\mathrm{dc}$ which has matrix elements up to $N=1$ given by
\begin{multline}
\bra{N, M_N} H \ket{N', M'_N} = \\
\begin{blockarray}{cccccc}
 \bra{0,0} & \bra{1,+1} & \bra{1,0} & \bra{1,-1} &\\
\begin{block}{(cccc)cc}
   0 & 0 & -d E / \sqrt{3} & 0 & & \ket{0,0} \\
   0 & 2 B_v & 0 & 0 & & \ket{1,+1}\\
   -d E / \sqrt{3} & 0 & 2 B_v & 0 & & \ket{1, 0}\\
   0 & 0 & 0 & 2 B_v & & \ket{1,-1}\\
\end{block}
\end{blockarray}
\label{eq:dcMatrix}
\end{multline}
where we have explicitly labelled the elements by $N, M_N, N', M'_N$ for clarity. The diagonal elements correspond to the rotational structure of $H_\mathrm{rot}$ and the off-diagonal terms from $H_\mathrm{dc}$ connect the $(N=0, M_N=0)$ and $(1,0)$ states with magnitude proportional to the electric field. This term mixes the two states, and effectively stops $N$ from being a good quantum number when an electric field is applied. We note that there are similar matrix elements that mix $N=1$ and $N=2$, and $N=2$ and $N=3$, and so on. Therefore, diagonalising Eq.~\ref{eq:dcMatrix} will not yield accurate energies for $N=0$ and $N=1$ in the presence of an electric field; rather the matrix must be extended to higher N to the point that the energies converge. 

\begin{figure}[t!]
\centering
\includegraphics[width=\columnwidth]{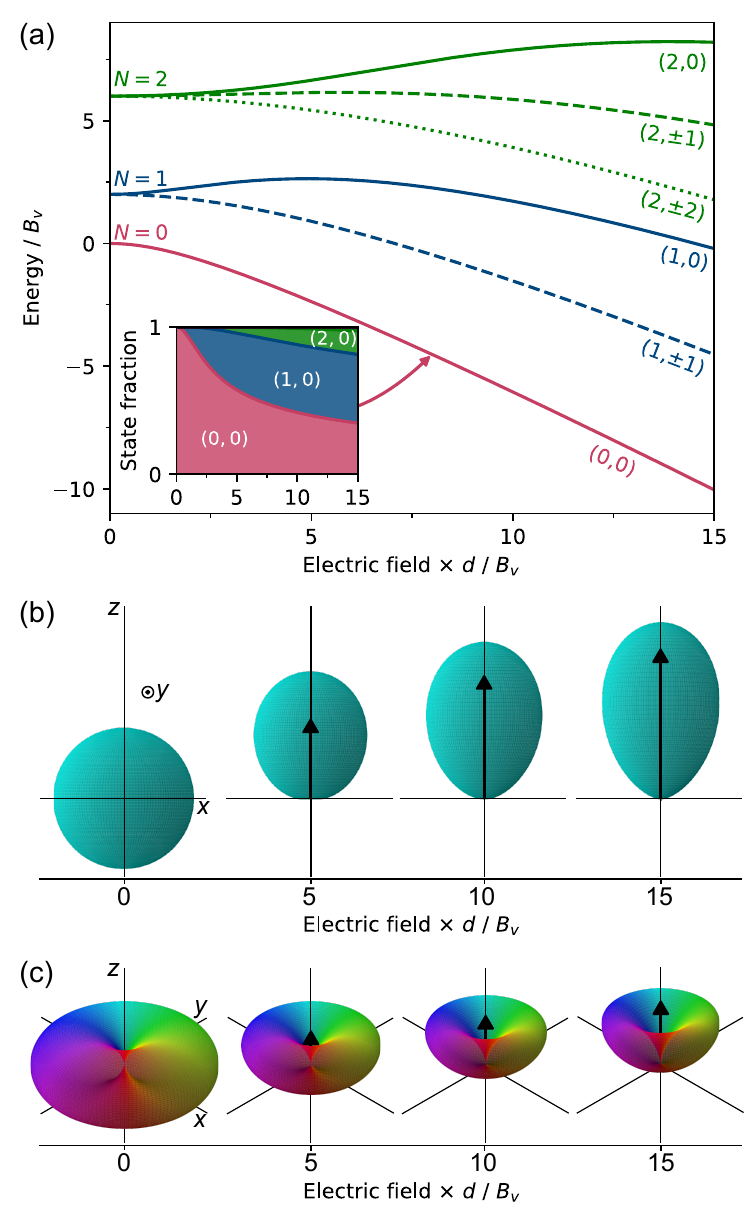}
\caption{Rigid-rotor molecule in an electric field. (a)~Shifts in energy as a function of electric field. At zero electric field, $N$ is a good quantum number and the $M_N$ sublevels of a given $N$ are all degenerate in energy. Application of an electric field couples neighbouring $N$ states that share the same $M_N$ resulting in mixing in the $N,M_N$ basis. We label the states at high field by $(\tilde{N}, M_N)$ where $\tilde{N}$ is the value of $N$ to which the state smoothly connects. The composition of the lowest energy state in the $(N, M_N)$ basis as a function of electric field is shown in the inset. The wavefunctions for the (b)~lowest energy state $({0},0)$ and (c)~second-lowest energy state $({1},\pm1)$ are illustrated at various electric fields, with the arrow indicating the direction and relative magnitude of the dipole moment.}
\label{fig:ElectricField}
\end{figure}

\begin{figure}[t!]
\centering
\includegraphics[width=0.99\columnwidth]{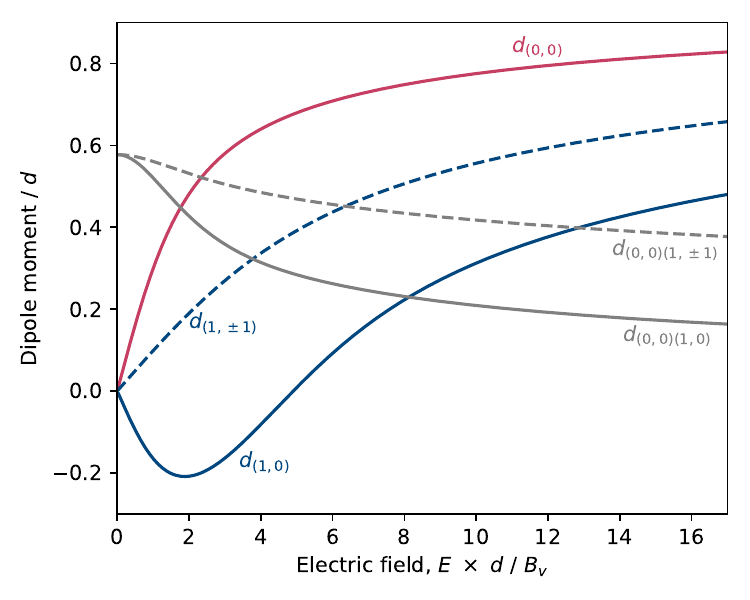}
\caption{Electric dipole moments associated with the rigid rotor molecule as a function of applied electric field. The red and blue lines show the static lab-frame dipoles associated with the states $d_{\tilde{N},M_N}$, equivalent to the gradient of the corresponding lines in Fig.~\ref{fig:RigidRotor}(a). The grey lines show the transition dipole moments between the states $d_{\tilde{N},M_N)(\tilde{N'},M'_N)}$. }
\label{fig:DipoleMoments}
\end{figure}

The result of this mixing is shown in Fig.~\ref{fig:ElectricField}. By diagonalising the Hamiltonian for $N_{\rm{Max}}=10$ we find both the state energies (the eigenvalues) \emph{and} the state composition in the $(N,M_N)$ basis (the eigenstates). In Fig.~\ref{fig:ElectricField}(a) we show the energy level structure of the rigid rotor molecule as a function of the magnitude of the electric field; the scales here are normalised by the quantities $B_v$ and $d$ such that they are valid for any hyperfine-free diatomic molecule. At zero electric field, we see the anharmonic ladder of rotational states labelled by $N$, while at large electric field the states are labelled by $\tilde{N}, M_N$, where $\tilde{N}$ is simply the rotational state that connects to that state if you follow it back to zero field. Note that $\tilde{N}$ is used to reflect the fact that the states become mixed in the $(N,M_N)$ basis. When the electric field is increased the coupling between $(N=0, M_N=0)$ and $(1,0)$, along with couplings to higher $N$, become stronger. The couplings cause the states to effectively repel each other. This repulsion allows for some intuition for how the states shift in energy as a function of the electric field. For example, $(\tilde{N}=0, M_N=0)$ shifts down in energy as it is repelled by $(1,0)$ above. In contrast, $(\tilde{N}=1, M_N=0)$ initially goes up in energy as repulsion from $(0,0)$ dominates, but goes down in energy at higher electric fields as repulsion from $(2,0)$ dominates. The lab-frame dipole of a given state is simply given by the gradient of each of the lines in Fig.~\ref{fig:ElectricField}(a), and can be evaluated by taking the derivative of the state energy with respect to the electric field. The dipole moments for the $N=0$~and~$1$ states are plotted in Fig.~\ref{fig:DipoleMoments}.

Evaluation of the wavefunctions gives a complete picture. The inset to Fig.~\ref{fig:ElectricField}(a) shows how the composition of the ground state, defined by the square of the coefficients of the eigenstates, changes as a function of electric field. At zero field, the state is completely described by $(N=0, M_N=0)$. However, as the field is increased, the state becomes an admixture of the $M_N=0$ component of various $N$ indicating that while $M_N$ remains a good quantum number for all but zero electric field $N$ does not. Using the eigenstate coefficients, we evaluate the probability distribution for the angle of the rigid rotor with respect to the electric field; these are shown for the states $(\tilde{N}=0, M_N=0)$ and $(1,\pm1)$ in Fig.~\ref{fig:ElectricField}(b,c). For both states shown, the wavefunction becomes asymmetric about the $xy$~plane such that the molecules become more likely to point in a preferred direction along the electric field direction ($z$ axis). This results in the lab-frame dipole moment indicated by the black arrow. Note that some states in Fig.~\ref{fig:ElectricField}(a), such as $(1,0)$, shift up in energy at low field but then shift down in energy at high field indicating that the direction of the dipole changes direction as the electric field is increased.

While a dc electric field naturally mixes the rotational states of the molecules, dipole moments in the lab frame can also be realised by preparing the molecules in quantum superpositions of different rotational states. For the simple rigid rotor molecule, electric dipole transitions may be resonantly driven between different rotational states following the selection rules for the rotational angular momentum $\Delta N = \pm1$, and $\Delta M_N=0, \pm1$ depending on the polarisation of the driving field. The rotational states are typically spaced in energy such that the required driving field is in the microwave frequency domain. When a transition is driven from an initially populated state to an unpopulated state, the  population periodically transfers back and forth between the two states in a Rabi cycle. By carefully selecting the duration for which the driving field is applied, an arbitrary superposition of the two states may be engineered. Note that, as transitions between rotational states are in the microwave domain, spontaneous emission is negligible. This makes the states very long-lived, and facilitates high fidelity Rabi oscillations.

\begin{figure*}[t!]
\centering
\includegraphics[width=2\columnwidth]{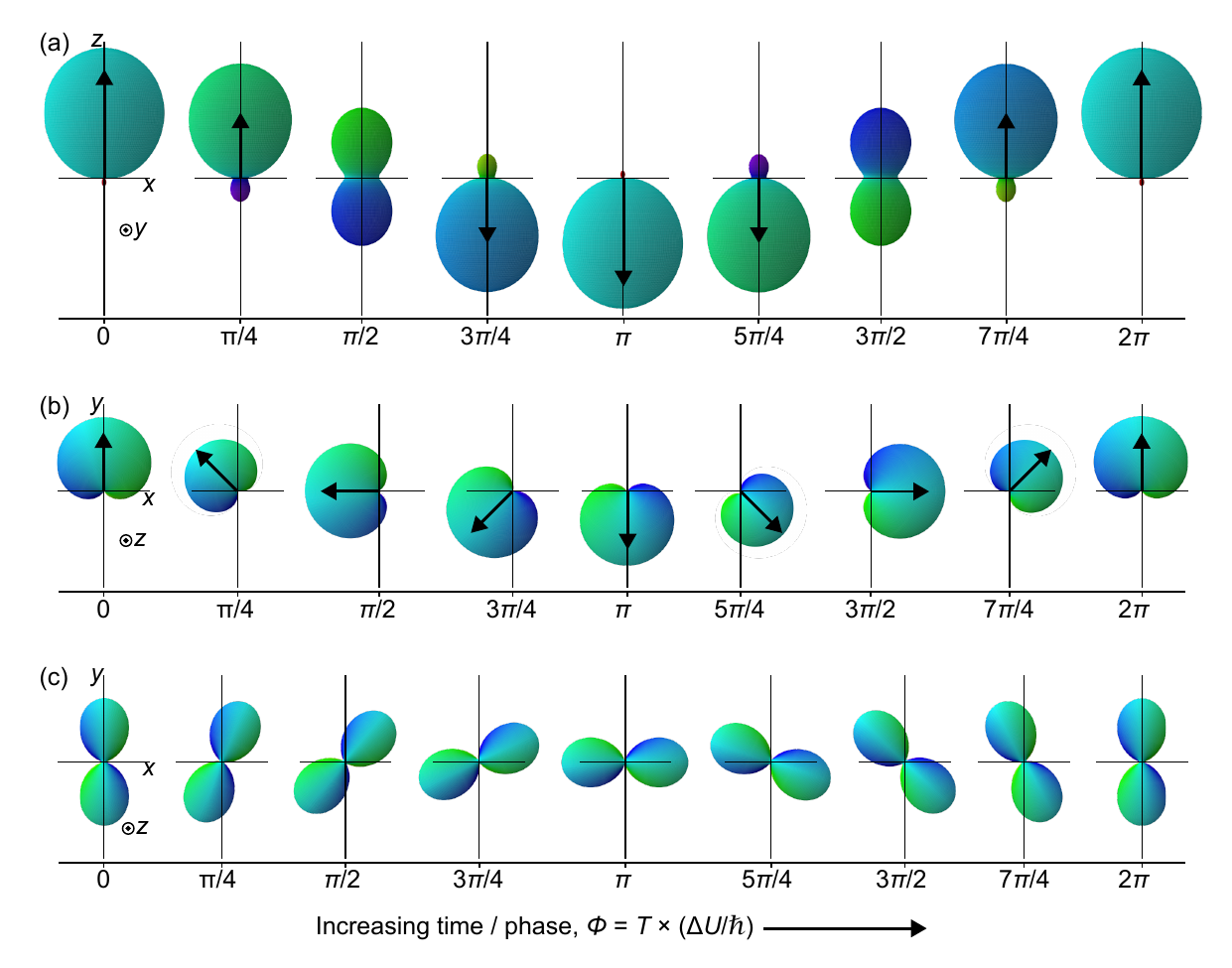}
\caption{Time and phase evolution for equal superpositions of the rigid-rotor wavefunctions \mbox{(a)~$(\ket{0,0}+\mathrm{e}^{-i\Phi}\ket{1,0})/\sqrt{2}$,} \mbox{(b)~$(\ket{0,0}+\mathrm{e}^{-i\Phi}\ket{1,1})/\sqrt{2}$,} \mbox{(c)~$(\ket{0,0}+\mathrm{e}^{-i\Phi}\ket{2,2})/\sqrt{2}$.} In each case, the direction and relative magnitude of the dipole is indicated by the black arrow. The evolution time $T$ is related to the phase of the superposition $\Phi=T\times(\Delta U/\hbar)$ where $\Delta U$ is the energy difference between the two superposed states and $\hbar$ is the reduced planck constant. }
\label{fig:OscillatingDipoles}
\end{figure*}

The superposition of two different rotational states does not result in a static dipole, but instead results in a dipole that oscillates at a frequency $\Delta U/h$ where $\Delta U$ is the difference in energy between the superposed states and $h$ is the Planck constant. This is due to the accumulation of a phase $\Phi = T \times \Delta U / h$ between the two components as the time $T$ evolves. The time-evolution of the angular probability distribution for 3 different superpositions of states is shown in Fig.~\ref{fig:OscillatingDipoles}. Note that here the calculations are performed in the absence of an electric field, although we still define the quantisation axis to be along the $z$ axis.  

For a pair of states connected by a dipole-allowed $\pi$-transition, the dipole oscillates along $z$. This is shown for the superposition of $(N=0, M_N=0)$ and $(1,0)$ states in Fig.~\ref{fig:OscillatingDipoles}(a). Here, the dipole flips periodically between pointing up and down, leading to on average the dipole for the superposition being zero. However, for dipoles oscillating in sync with each other a non-zero dipole-dipole interaction will be present. Take, for example, two synchronised molecules separated by a distance along $x$. When $\Phi=0$ the two dipoles are parallel side-by-side and so experience a repulsive interaction. When $\Phi=\pi/2$ the dipoles are fully oriented in the opposite direction, but the two side-by-side dipoles are still parallel to one another and therefore will still experience a repulsive interaction.

The magnitude of the dipole that contributes to the interaction is the transition dipole moment that connects the two states. This is described by the dipole operator that has matrix elements given by
\begin{multline}
\bra{N,M_M}\hat{d}\ket{N',M'_N} = \\
-d \sqrt{(2N+1)(2N'+1)} (-1)^{M_N} \\
\times \begin{pmatrix}
N & 1 & N'\\
-M_N & M & M'_N
\end{pmatrix}
\begin{pmatrix}
N & 1 & N'\\
0 & 0 & 0
\end{pmatrix}.
\label{eq:dipMatrix}
\end{multline}
This is similar to Eq.~\ref{eq:dcMatrixElements} only without the dependence on electric field and with the addition of the variable $M$. The properties of the Wigner-3j symbols result in Eq.~\ref{eq:dipMatrix} being non-zero only when $M=M_N-M'_N=0,\pm1$. For $M=0$ the transition dipole corresponds to excitation along the quantisation ($z$) axis such that the transition can be driven with the microwaves polarised parallel to this axis. Meanwhile $M=\pm1$ transitions corresponds to excitation perpendicular to the quantisation axis (in the $xy$~plane), and are associated with $\sigma^\pm$ transitions. 

The time evolution for the superposition of $(N=0, M_N=0)$ and $(1,1)$, relating to a $\sigma^+$ transition, is shown in Fig.~\ref{fig:OscillatingDipoles}(b). Here we see that the magnitude of the dipole remains constant, and rotates in the $xy$~plane around the $z$~axis. As with the earlier superposition, a similar picture can be built with two molecules side-by-side separated along $x$. However this time the dipoles will oscillate between parallel and anti-parallel configurations, leading to weaker interactions than those associated with the $\pi$ transition. As the direction of oscillation is defined by $z$, this remains the relevant direction for defining $\Theta$ in the geometry of the dipole-dipole interactions. However, with the dipole restricted to the $xy$~plane the strength of dipole-dipole interactions for oscillating dipoles from the $\sigma$ transitions are related to those of an equivalent $\pi$ transition by a factor of $-1/2$, i.e. they are weaker and have the opposite sign. 

The strength of the transition dipole moments may also be modified by dc electric fields. This is due to the aforementioned mixing of states in the $N$ basis. The transition dipole moments for the transitions between $N=0$~and~$1$ are shown by the grey lines in Fig.~\ref{fig:DipoleMoments}. Note that at zero electric field the transition dipole moments are $d/\sqrt{3}$, making the use of oscillating dipoles any easy approach to generate sizeable dipolar interactions.

Finally, we examine what happens when we prepare a superposition of two states that are not directly connected via an electric dipole-allowed transition. The angular probability distribution for the superposition of $(N=0, M_N=0)$ and $(2,2)$ is shown in Fig.~\ref{fig:OscillatingDipoles}(c). These states violate the selection rule $\Delta N = \pm 1$ and as such have no dipole moment association with the transition between the states. In this case we find that the probability distribution still shows a familiar rotating behaviour, but there is a symmetry to the distribution that results in a net-zero dipole moment at all times. 

Finally we note that such oscillating dipoles may be accessed in different ways. For example, the dressed states associated with a molecule in a continuous but off-resonant driving field are also equivalent to free-space superpositions of states. In section~\ref{sec:shielding}, we will briefly discuss microwave shielding which is a powerful technique for controlling collisional loss of molecules, and exploits the interactions between oscillating dipoles that result from off-resonant dressing on rotational transitions. Recent advances in double-microwave shielding show how these interactions may be made highly tunable by combining vertically-oscillating dipoles with dipoles rotating in the horizontal plane~\cite{Karman2025}.

\subsection*{Dipole-dipole interactions and spin exchange}

The general form for the interaction energy between two neighbouring dipoles is described classically by 
\begin{equation}
V_{\rm{dd}}=\frac{1}{4\pi\epsilon_0} \frac{\vec{d_i}.\vec{d_j}-3(\vec{d_i}.\vec{e_r})(\vec{d_j}.\vec{e_r})}{r^3},
\end{equation}
where $\vec{e_r}$ is a unit vector connecting the two dipoles and $r$ is the distance between them. This simplifies to Eq.~\ref{eq:DipoleDipole} for the case where the dipoles have the same orientation. The quantum mechanical form of this equation requires rewriting in terms of operators. More advanced texts~\cite{Wall_2015b} show that the dipolar interaction is the contraction of two rank-two tensors, one acting on the internal state of the molecules through 
the dipole operators and one acting on the orbital motion of the two molecules through the angular dependence. It follows that the dipolar interaction can be written as a sum of terms which transfer $q$ units of rotational angular momentum
projection to orbital angular momentum projection with $-2 \leq q \leq 2$. In polar coordinates, the Hamiltonian that describes the interaction between two molecules labelled $i$~and~$j$ is then~\cite{Wall_2015b}
\begin{equation}
\hat{H}_{\mathrm{dd}}=\hat{H}^{q=0}_{\mathrm{dd}}+\hat{H}^{q=\pm1}_{\mathrm{dd}}+\hat{H}^{q=\pm2}_{\mathrm{dd}},
\end{equation}
where
\begin{multline}
\hat{H}^{q=0}_{\mathrm{dd}} = \frac{1-3\cos^2\Theta}{4\pi\epsilon_0 r^3} \\
\times \left(  \hat{d}^{(i)}_0 \hat{d}^{(j)}_0 + \frac{\hat{d}^{(i)}_{-1} \hat{d}^{(j)}_{+1} + \hat{d}^{(i)}_{+1} \hat{d}^{(j)}_{-1}}{2}  \right),
\label{eq:Hdd0}
\end{multline}
\begin{multline}
\hat{H}^{q=\pm1}_{\mathrm{dd}} = - \frac{3}{\sqrt{2}} \frac{\sin\Theta \cos\Theta}{4\pi\epsilon_0 r^3} \\
\times \left[e^{+i\phi}\left(\hat d_0^{(i)}\hat d_{-1}^{(j)} + \hat d_{-1}^{(i)}\hat d_0^{(j)} \right)+\mathrm{h.c.}\right],
\label{eq:Hdd1}
\end{multline}

\begin{equation}
\hat{H}^{q=\pm2}_{\mathrm{dd}} = - \frac{3}{2}\frac{\sin^2\Theta}{4\pi\epsilon_0 r^3}
\left(e^{+2i\phi}\hat d_{-1}^{(i)}\hat d_{-1}^{(j)} + \mathrm{h.c.} \right).
\label{eq:Hdd2}
\end{equation}
Here $\hat{d}_0, \hat{d}_{-1}, \hat{d}_{+1}$ are spherical components of the dipole operator. The terms in Eqs.~\ref{eq:Hdd1} and~\ref{eq:Hdd2} do not separately conserve internal and external angular momentum; they transfer $q$ units of angular momentum from molecular rotation to orbital motion and vice versa. As we shall see, hyperfine and Zeeman interactions lift the degeneracy of $M_N$ states such that these terms are usually far from resonance and therefore are negligible for molecules. However, with careful experimental design such terms may be brought to resonance, and, for example, the $q=\pm2$ term can be exploited to realise spin-orbit coupling with ultracold molecules~\cite{Syzranov2014}.

For the purposes of the discussion here, we therefore confine our attention to the $q=0$ term in Eq.~\ref{eq:Hdd0} which conserves the internal and orbital angular momentum separately. The first term in Eq.~\ref{eq:Hdd0} corresponds to dipoles oriented along $z$ such as those realised by applying a dc electric field or a transition dipole associated with a $\pi$ transition. The second term corresponds to dipoles associated with $\sigma$ transitions with the factor of $1/2$ formalising the weaker interactions arising from the geometry. 

\begin{figure}[t!]
\centering
\includegraphics[width=\columnwidth]{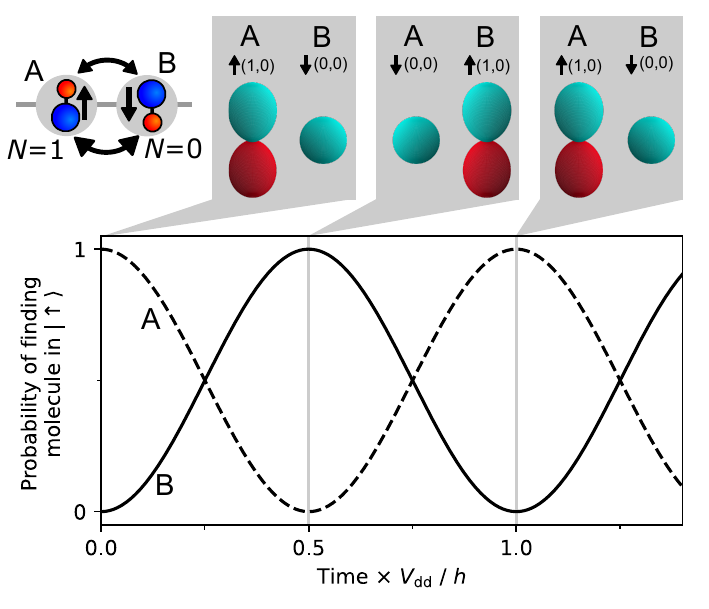}
\caption{Dipolar spin exchange between two molecules that are initially prepared in different rotational states. For molecules prepared in states connected by a non-zero transition dipole moment there exists an interaction with energy $V_\mathrm{dd}\propto d_\mathrm{\uparrow\downarrow}^2/r^3$. This leads to the periodic exchange of rotational energy between the molecules. This is illustrated with the two molecules above labelled A, B. We prepare the molecules such that A is in $\ket{\uparrow}\equiv(N=1, M_N=0)$ and B is in $\ket{\downarrow}\equiv(0,0)$. At a later time $V_\mathrm{dd}/(2h)$ this situation becomes reversed with A in $\ket{\downarrow}$ and B in $\ket{\uparrow}$. Between these times the two molecules enter an entangled superposition of the two spin states and each possess an oscillating lab-frame dipole as illustrated in Fig.~\ref{fig:OscillatingDipoles}(a). Dipolar spin exchange forms the basis of an entangling gate protocol for quantum computing with polar molecules as shown in section~\ref{sec:QuantumSimulation}. }
\label{fig:SpinExchange}
\end{figure}

To study the matrix elements of Eq.~\ref{eq:Hdd0} we use a spin mapping notation, such that two rotational states are given the labels of a spin-1/2 basis~$\{ \uparrow, \downarrow \}$. Here we choose the spin mapping $\ket{\downarrow}=\ket{N=0, M_N=0}$ and $\ket{\uparrow}=\ket{N=1, M_N=0}$. Expanding that basis across the two molecules yields four possible configurations resulting in a general set of Hamiltonian matrix elements
\begin{multline}
\bra{N, M_N} \hat{H}_\mathrm{dd} \ket{N', M'_N} = \frac{1-3\cos^2\Theta}{4\pi\epsilon_0 r^3} \times \\
\begin{blockarray}{cccccc}
 \bra{\downarrow^{(i)}\downarrow^{(j)}} & \bra{\downarrow^{(i)}\uparrow^{(j)}} & \bra{\uparrow^{(i)}\downarrow^{(j)}} & \bra{\uparrow^{(i)}\uparrow^{(j)}} &\\
\begin{block}{(cccc)cc}
   d^2_\downarrow & 0 & 0 & 0 & & \ket{\downarrow^{(i)}\downarrow^{(j)}} \\
   0 & d_\downarrow d_\uparrow & d^2_{\uparrow\downarrow} & 0 & & \ket{\downarrow^{(i)}\uparrow^{(j)}}\\
   0 & d^2_{\uparrow\downarrow} & d_\downarrow d_\uparrow & 0 & & \ket{\uparrow^{(i)}\downarrow^{(j)}}\\
   0 & 0 & 0 & d^2_\uparrow & & \ket{\uparrow^{(i)}\uparrow^{(j)}}\\
\end{block}
\end{blockarray}
\label{eq:Vdd_Ham_Matrix}
\end{multline}
where $d_\downarrow, d_\uparrow$ correspond to the expectation value for the static dipole moment of the states $\ket{\downarrow}, \ket{\uparrow}$ respectively, and $d_{\uparrow\downarrow}$ represents the transition dipole matrix element that connects the two states. These off-diagonal elements are responsible for interactions between the oscillating dipoles, and also leads to flipping between the $\downarrow^{(i)}\uparrow^{(j)}$ and $\uparrow^{(i)}\downarrow^{(j)}$ configurations as shown in Fig.~\ref{fig:SpinExchange}. This transfer of rotational angular momentum between pairs of molecules is commonly known as spin exchange and can be used to controllably engineer quantum entanglement~\cite{Ni2018}. 

Finally, a useful reformulation of Eq.~\ref{eq:Hdd0} can be performed by defining the spin-1/2 operators 
\begin{equation}
\hat{S}_z^{(i)}=\frac{1}{2}(\ket{\uparrow^{(i)}}\bra{\uparrow^{(i)}}- \ket{\downarrow^{(i)}}\bra{\downarrow^{(i)}}),
\end{equation}
\begin{equation}
\hat{S}_+^{(i)} = \ket{\uparrow^{(i)}}\bra{\downarrow^{(i)}},
\end{equation}
\begin{equation}
\hat{S}_-^{(i)} = \ket{\downarrow^{(i)}}\bra{\uparrow^{(i)}},
\end{equation}
leading to the Hamiltonian
\begin{multline}
\hat{H}^{q=0}_{\mathrm{dd}}=\frac{1-3\cos^2\Theta}{4\pi\epsilon_0 r^3} \times\\ 
\left( \frac{J_\perp}{2}(\hat{S}_+^{(i)}\hat{S}_-^{(j)} + \hat{S}_-^{(i)}\hat{S}_+^{(j)}) + J_z \hat{S}_z^{(i)}\hat{S}_z^{(j)}  \right).
\label{eq:SpinHamiltonian}
\end{multline}
This separates the two mechanisms with the term $J_\perp=2d^2_{\uparrow\downarrow}$ describing the rate of dipolar spin exchange, while $J_z=(d_\uparrow-d_\downarrow)^2$ just describes the effects from static dipoles that arise from applied dc electric fields. This Hamiltonian describes the well-known XXZ spin model from quantum magnetism. The nomenclature is transparent when one realises that $\hat{S}_+^{(i)}\hat{S}_-^{(j)}+\hat{S}_-^{(i)}\hat{S}_+^{(j)}= 2(\hat{S}_x^{(i)}\hat{S}_x^{(j)}+\hat{S}_y^{(i)}\hat{S}_y^{(j)})$.

\begin{figure}[t!]
\centering
\includegraphics[width=\columnwidth]{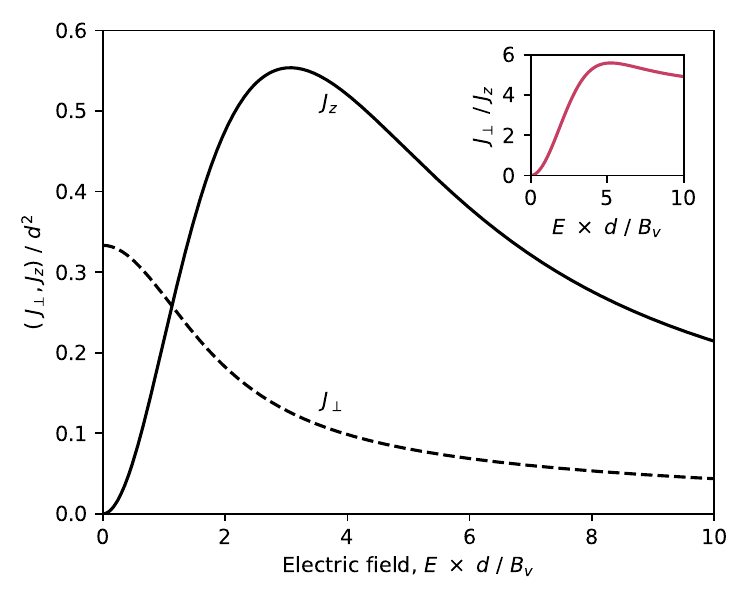}
\caption{Comparison of the relative magnitudes of the terms $J_\perp$ and $J_z$.
as a function of electric field for the rigid-rotor state combination~$(N=0,M_N=0)+(1,0)$. The inset shows how the ratio of these terms varies.}
\label{fig:J}
\end{figure}

The relative strengths of $J_\perp$ and $J_z$ vary as a function of electric field as shown in Fig.~\ref{fig:J}. Each of the plotted lines being derived from the dipole moments in Fig.~\ref{fig:DipoleMoments}. Note, that this plot depends strongly upon the spin mapping scheme that is chosen. For the chosen mapping of $\ket{\downarrow}=\ket{0,0}, \ket{\uparrow}=\ket{1,0}$, we see that the $J_z$ term increases to a maximum at an applied electric field around $\sim3d/B_v$, while the $J_\perp$ term reduces in strength. The ratio of these terms is also important as phase transitions may occur as the relative strengths of competing interactions is varied. This ratio is shown inset, and reaches a maximum value of 5.6 for this state mapping at an electric field of $\sim5d/B_v$. At an electric field of $\sim1.1d/B_v$ the values of $J_\perp$ and $J_z$ are equal, i.e.~$J_\perp=J_z$, which realises the Heisenberg spin model~\cite{Wall2015}.

\subsection*{Introducing nuclear spin}

So far we have considered the diatomic molecule as only possessing rotational angular momentum. However, the component nuclei typically also possess non-zero nuclear spin that gives rise to rich hyperfine structure. Moreover, interactions between the nuclear angular momentum and rotational angular momentum can lead to further mixing and complexity. 

Adding nuclear spin introduces three additional hyperfine interactions to the Hamiltonian for bialkli molecules
\begin{equation}
H_\mathrm{hf} = H_\mathrm{quad} + H_\mathrm{spin-spin} + H_\mathrm{spin-rotation}.
\end{equation}
Here $H_\mathrm{quad}$ corresponds to the nuclear quadrupole interaction between the nuclear quadrupole moment and the electric field gradient due to the distribution of electrons. $H_\mathrm{spin-spin}$ describes the interaction between the two nuclear spins, and $H_\mathrm{spin-rotation}$ describes the interaction between the nuclear spin and the rotational angular momentum of the molecule. These terms are fully defined in~\cite{Blackmore2023}. 

To illustrate the effects of nuclear spin we will use $^{87}$Rb$^{133}$Cs (hereafter RbCs) as our main example molecule. This molecule has been used throughout the author's experiments and is particularly relevant to the concept of magic-wavelength traps discussed later. In general, hyperfine structures are similar across all bialkali molecules, though with sometimes large changes in relative energies and coupling strengths. For those interested, similar plots to those shown here  are given in~\cite{Blackmore2023} for a range of bialkali molecules, along with open-source Python codes to perform calculations for other singlet molecules.   

\begin{figure*}[t!]
\centering
\includegraphics[width=2\columnwidth]{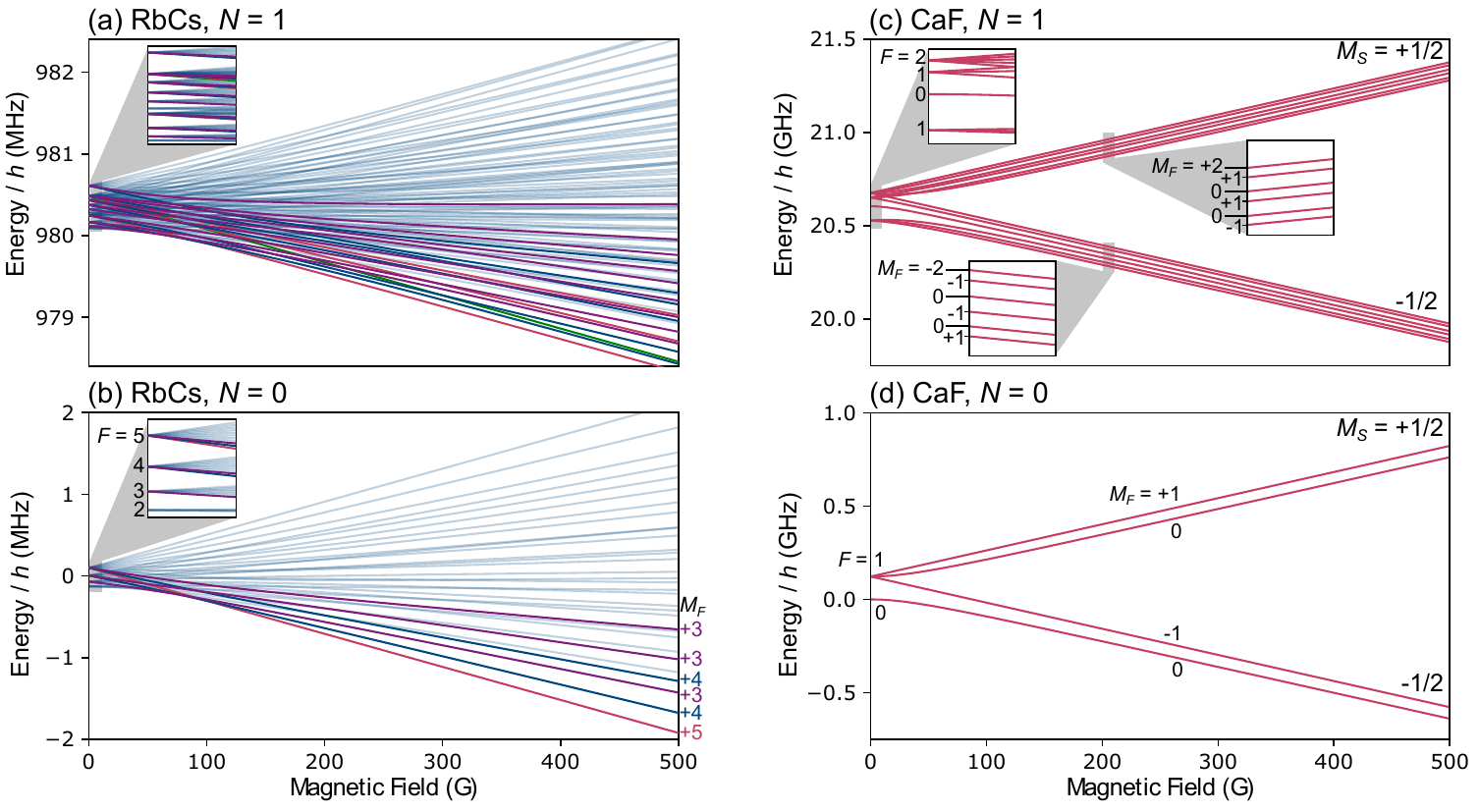}
\caption{Comparison of the rotational and hyperfine structure in a magnetic field for the RbCs molecule in (a) $N=1$ and (b) $N=0$ in the $^1\mathrm{\Sigma}$ ground vibrational state and the CaF molecule in (c) $N=1$ and (d) $N=0$ in the $^2\Sigma$ ground vibrational state. For RbCs $I_\mathrm{Rb}=3/2$, $I_\mathrm{Cs}=7/2$, and for CaF $I_\mathrm{Ca}=0$, $I_\mathrm{F}=1/2$. Due to the dense hyperfine structure in RbCs, we have highlighted and colour-coded states only with $M_F=3,4,5,6$. Labels for $M_F$ are given for $N=0$, the same colour coding applied for the $N=1$ panel, with the additional green line showing the spin-stretched $M_F=6$ state. }
\label{fig:ComparingMolecules}
\end{figure*}

The hyperfine structure of the ground and first-excited rotational states of RbCs is shown as a function of magnetic field in Fig.~\ref{fig:ComparingMolecules}(a,b). The magnetic field couples to both the nuclear spins and the rotational angular momentum due to the Zeeman Hamiltonian
\begin{equation}
H_\mathrm{Zeeman} = -g_r\mu_\mathrm{N}\boldsymbol{N}\cdot\boldsymbol{B} - \sum_{j=\mathrm{Rb,Cs}}g_j(1-\sigma_j)\mu_\mathrm{N}\boldsymbol{I}\cdot\boldsymbol{B}.
\label{eq:ZeemanHam}
\end{equation}
The first term corresponds to the Zeeman shift associated with the interaction between the rotational angular momentum of the molecule and the magnetic field, which is scaled by the factor~$g_r$. The second term corresponds to the Zeeman shift associated with each of the nuclear spins, which are scaled by the nuclear spin $g_j$-factors and incorporates an isotropic shielding $\sigma_j$. The term $\mu_\mathrm{N}$ is the nuclear magneton. For bialkali molecules $g_r \ll g_j$ such that the Zeeman shifts are dominated by the interaction with the nuclear spins. As the Zeeman shifts associated with nuclear spins are still relatively small, the many hyperfine states available in bialkali molecules are densely packed into a small \mbox{$\sim1$\,MHz} energy range. 

At zero field, the total angular momentum $F=I_\mathrm{Rb}+I_\mathrm{Cs}+N$ is a good quantum number. In $N=0$ the nuclear spins of $I_\mathrm{Rb}=3/2$, $I_\mathrm{Cs}=7/2$ lead to values of $F$ in the range $|I_\mathrm{Cs}-I_\mathrm{Rb}|<F<|I_\mathrm{Cs}+I_\mathrm{Rb}|$ such that there are four hyperfine levels with $F=2,3,4,5$. When a magnetic field is applied, these states split into $(2I_\mathrm{Rb}+1)(2I_\mathrm{Cs}+1)(2N+1)$ Zeeman sublevels, such that there are 32 hyperfine states in $N=0$ and 96 hyperfine states in $N=1$. At large magnetic field these states become well described by the uncoupled basis $(N, M_N, m_\mathrm{Rb}, m_\mathrm{Cs})$, where $m_\mathrm{Rb},m_\mathrm{Cs}$ are the projections of the nuclear spins $I_\mathrm{Rb},I_\mathrm{Cs}$ along the quantisation axis. However for many molecules, including RbCs, this asymptotic limit is not reached until very large magnetic fields are applied, typically $\gtrsim1000$\,G. Experiments often operate in an intermediate regime where the only good quantum numbers are $N$ and the total angular momentum projection $M_F=M_N+m_\mathrm{Rb}+m_\mathrm{Cs}$. As these quantum numbers are not enough to uniquely identify the majority of states, we have adopted the notation $(N, M_F)_k$ where $k$ is an index that counts the states with a given $N, M_F$ in order of ascending energy. In Fig.~\ref{fig:ComparingMolecules}(a,b) we have highlighted and colour-coded states with $M_F=3,4,5,6$. 

To contrast with the bialkali molecule RbCs, we show the equivalent Zeeman structure associated with the laser-cooled CaF molecule in Fig.~\ref{fig:ComparingMolecules}(c,d). The electronic ground states of these molecules are different. RbCs has a $^{1}\Sigma$ ground state that has electronic spin $S=0$, indicating that the free electrons contributed from the component Rb and Cs atoms (one from each) are antialigned. CaF on the other hand has a $^2\Sigma$ ground state with one unpaired electron leading to an electronic spin $S=1/2$. This additional spin must be incoporated into the total angular momentum $F=I_\mathrm{Ca}+I_\mathrm{F}+N+S$, and importantly the electron spin can interact with the magnetic field via an additional Zeeman term $g_s \mu_\mathrm{B} \boldsymbol{S}\cdot\boldsymbol{B}$, where $\mu_\mathrm{B}$ is the Bohr magneton. This term is orders of magnitude larger than the terms in Eq.~\ref{eq:ZeemanHam}, and leads to Zeeman shifts of the order of \mbox{MHz/G}. Also note that the structure of CaF appears relatively simple compared to RbCs due to the much smaller nuclear spins of the component nuclei $I_\mathrm{Ca}=0, I_\mathrm{F}=1/2$. Open-source Python codes for calculating the structure of $^2\Sigma$ molecules can be found in~\cite{Humphreys2025}. 

\begin{figure}[t!]
\centering
\includegraphics[width=1.0\columnwidth]{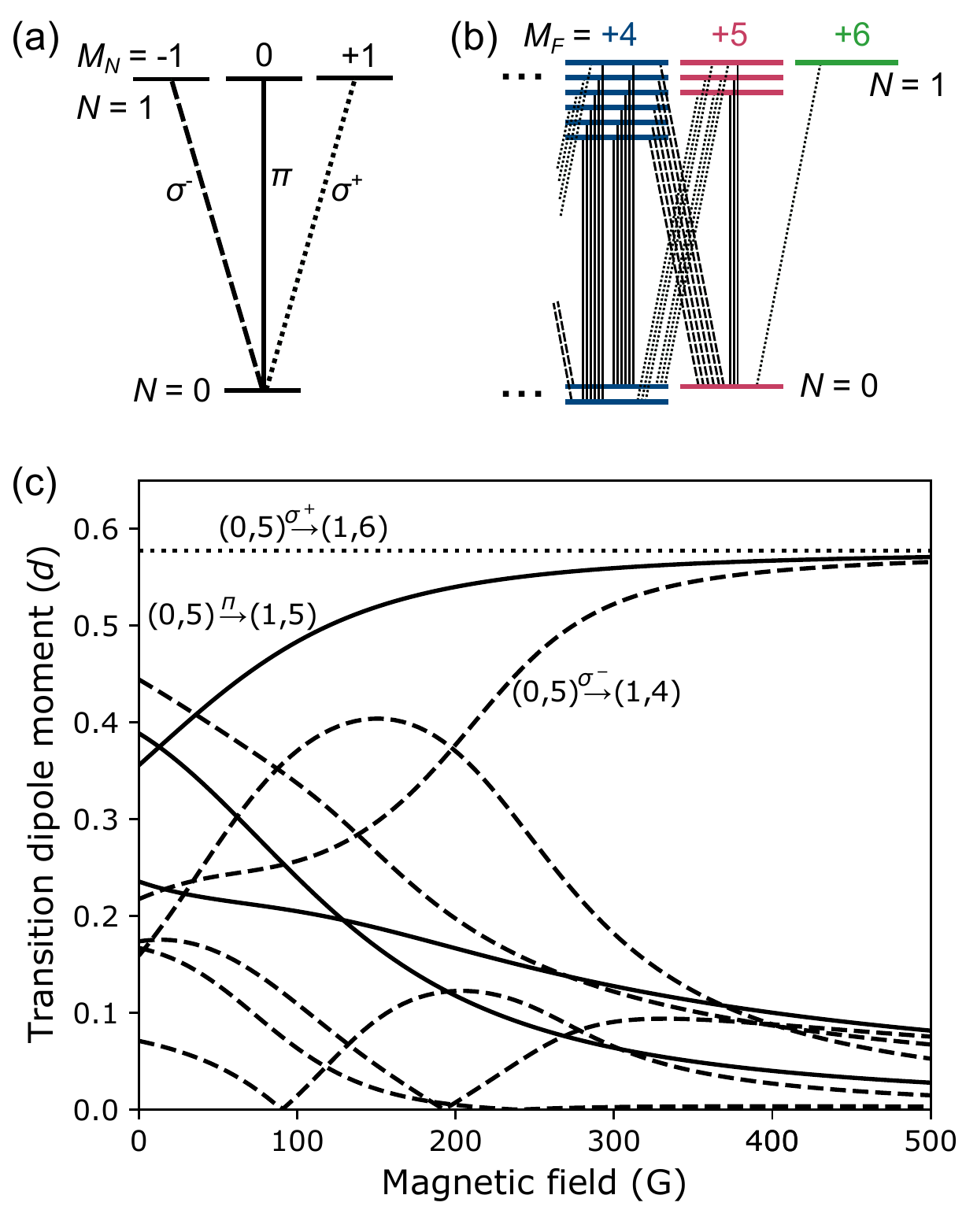}
\caption{Effect of nuclear spin on the number of available transitions between the $N=0$ and $N=1$ rotational states. (a)~For zero nuclear spin ($I=0$) there are 3 available transitions from $N=0$ which change $M_N$ by $0, \pm1$ depending on the polarisation of the driving field; the polarisation if the driving field is indicated by dashed, solid, and dotted lines throughout the figure. (b)~The situation for $^{87}$Rb$^{133}$Cs with $I_\mathrm{Rb}=3/2, I_\mathrm{Cs}=7/2$, where $M_F$ is a good quantum number instead of $M_N$. There is one transition between states that are spin stretched e.g. $(N=0, M_F=5)\rightarrow(N=1, M_F=6)$, but many more transitions allowed between non-stretched states with the magnitude of the transition dipole moment depending on the overlap between state components with the same nuclear spin. (c)~The transition dipole moments, in units of the molecule-frame dipole moment, for $^{87}$Rb$^{133}$Cs given for all available transitions from the $(N=0, M_F=5)$ state as a function of magnetic field. }
\label{fig:TransitionDipole}
\end{figure}

Driving transitions between rotational states is enriched by the addition of nuclear spin. The selection rules for rotational transitions are $\Delta N = \pm1$, and $\Delta M_N=0,\pm1$ depending on the polarisation of the driving field. Incorporating the nuclear spins, we can impose additional selection rules that the angular momentum projection associated with the nuclear spins cannot be changed; for RbCs this would mean $\Delta m_\mathrm{Rb}=0, \Delta m_\mathrm{Cs}=0$. However, as $M_N$ is not necessarily a good quantum number, the rules for the projected angular momentum are often propagated through to the total angular momentum projection to become $\Delta M_F=0,\pm1$. This greatly increases the number of allowed transitions, as illustrated in Fig.~\ref{fig:TransitionDipole}.

As a worked example, we consider RbCs molecules prepared in the state $(N=0, M_F=5)_0$ at a magnetic field of 181.5\,G. This state is spin stretched and so can be described completely in the uncoupled basis as 
\begin{equation*}
(0,5)_0=\ket{N=0, M_N=0, m_\mathrm{Rb}=3/2, m_\mathrm{Cs}=7/2}.
\end{equation*}
Applying microwaves polarised along the quantisation axis will drive $\pi$ transitions to states in $N=1$ with $M_F=5$. There are three such states
\begin{align*}
(1,5)_0 = & ~ 0.925\ket{1,0,3/2,7/2} \\ 
& + 0.087\ket{1,1,3/2,5/2} \\
& + 0.370\ket{1,1,1/2,7/2}, \\
\end{align*}
\begin{align*}
(1,5)_1 = & ~ -0.229\ket{1,0,3/2,7/2} \\ 
& + 0.905\ket{1,1,3/2,5/2} \\
& - 0.358\ket{1,1,1/2,7/2}, \\
\end{align*}
\begin{align*}
(1,5)_2 = & ~ -0.304\ket{1,0,3/2,7/2} \\ 
& + 0.416\ket{1,1,3/2,5/2} \\
& - 0.857\ket{1,1,1/2,7/2}. \\
\end{align*}
The relative strengths of the transitions to the above $M_F$ states are given by the overlap of the components that satisfy the electric dipole selection rules. That is, the transition strength is scaled by the coefficient of the $\ket{1,0,3/2,7/2}$ state component. By evaluating Eq.~\ref{eq:dipMatrix} with $M=0$, we can determine that the transition dipole matrix element that connects states with \mbox{$N=0, M_N=0$} and \mbox{$N'=1, M'_N=0$} is equal to $d/\sqrt{3}$. The transition dipoles that connect the available states are therefore
\begin{equation*}
(0,5)_0 \rightarrow (1,5)_0: 0.925\times d/\sqrt{3},
\end{equation*}
\begin{equation*}
(0,5)_0 \rightarrow (1,5)_1: 0.229\times d/\sqrt{3},
\end{equation*}
\begin{equation*}
(0,5)_0 \rightarrow (1,5)_2: 0.304\times d/\sqrt{3}.
\end{equation*}
This can become more complicated when the total transition dipole must be summed over multiple state components that all match the relevant selection rules. 

The mixing of uncoupled basis states provides an indirect mechanism for electric dipole transitions to change the nuclear spin~\cite{Ospelkaus2010}. For example, once molecules are populated into one of the $N=1, M_F=5$ states they possess the character of three different combinations of nuclear spins. A second microwave, this time circularly polarised such that it drives $\sigma^+$ transitions can be used to drive from the components with nuclear spin projections $(m_\mathrm{Rb}=3/2, m_\mathrm{Cs}=5/2)$ and $(1/2, 7/2)$ to one of the two $N=0, M_F=4$ states that have composition
\begin{align*}
(0,4)_0 = & ~ 0.947\ket{1,0,3/2,5/2} \\ 
& + 0.321\ket{1,0,1/2,7/2}, \\
\end{align*}
\begin{align*}
(0,4)_1 = & ~ 0.321\ket{1,0,3/2,5/2} \\ 
& + 0.947\ket{1,0,1/2,7/2}. \\
\end{align*}
Therefore even though the electric dipole selection rules prohibit changes to the nuclear spin, it is still allowed by transferring the molecules via states with mixed nuclear spin character. The ability to transfer the molecules between states with different nuclear spin in this way unlocks the hyperfine degree of freedom as an additional resource for quantum science. For example hyperfine states in the rotational ground state, differing only in their nuclear spin composition, may be used as a robust quantum memory~\cite{Park2017, Gregory2021, Lin2022}. 

The composition of the states, and therefore the transition dipole moment, changes with the applied magnetic field. We show how all of the transitions from the $(0,5)_0=(N=0, M_N=0, m_\mathrm{Rb}=3/2,m_\mathrm{Cs}=7/2)$ state varies with magnetic field in Fig.~\ref{fig:TransitionDipole}(c), with $\sigma^-, \pi, \sigma^+$ transitions indicated by dashed, solid, and dotted lines respectively. 
We see that the spin-stretched $(0,5)_0\rightarrow(1,6)_0$ transition ($\sigma^+$, dotted line) has a transition dipole equal to $d/\sqrt{3}\approx 0.577 d$ at all magnetic fields. All of the other transitions on the other hand experience significant variation over the 500\,G field range shown. At large field, the nuclear spins begin to decouple such that $m_\mathrm{Rb}, m_\mathrm{Cs}$ becomes a good quantum number. This is indicated by the weakening of all but 3 transitions corresponding to states with increasingly dominant $m_\mathrm{Rb}=3/2, m_\mathrm{Cs}=7/2$ nuclear spin contributions.

\begin{figure}[t!]
\centering
\includegraphics[width=\columnwidth]{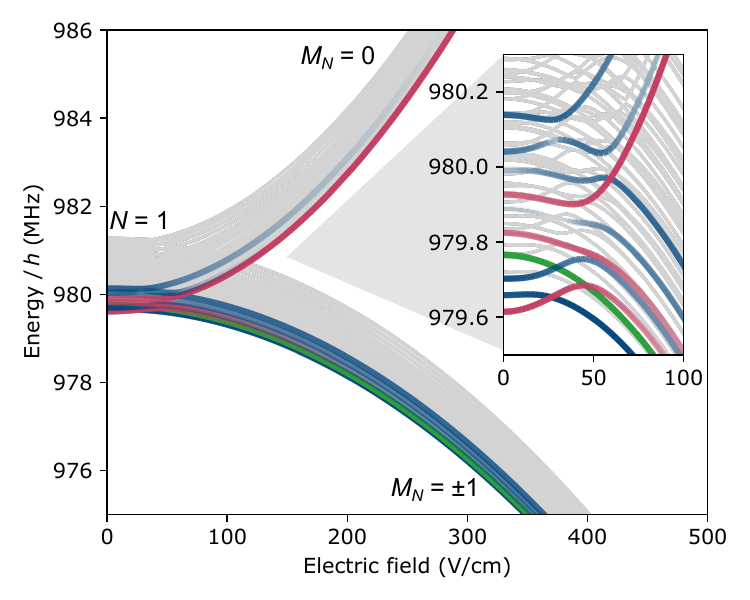}
\caption{Response of the $N=1$ state of RbCs in response to an applied dc electric field. The electric field is applied parallel to an additional 181.5\,G magnetic field. States are coloured by the relative strength of transitions from the spin-stretched ground state $(N=0,M_F=5)$, with blue, red, green indicating $\sigma^-, \pi, \sigma^+$ transitions to $N=1$ states with $M_F=4,5,6$ respectively. }
\label{fig:RbCs_electricfield}
\end{figure}

We now examine the effect of the electric field on the molecule including hyperfine structure. In Fig.~\ref{fig:RbCs_electricfield} we show the dc Stark shift for the RbCs hyperfine states in $N=1$. In addition to the electric field, a magnetic field of 181.5\,G is also present in the calculation to lift the degeneracy between $M_F$ levels. This time we colour code the states by the strength of transitions from the $(0,5)_0$ state, such that the colours indicate states with $M_F=4,5,6$ and the intensity indicates the fraction of that state with nuclear spin character $m_\mathrm{Rb}=3/2, m_\mathrm{Cs}=7/2$. The behaviour broadly matches the simple rigid rotor model illustrated in Fig.~\ref{fig:ElectricField}. However, when the dc Stark shift is of similar magnitude to the hyperfine structure splitting, couplings between states lead to a complicated structure of avoided crossings and significant variation in the strengths of available transitions, as highlighted in the inset to Fig.~\ref{fig:ElectricField}. Note that the electric field again only couples to the rotational angular momentum $N, M_N$ and does not couple directly to the nuclear spin. 

\section{Optical trapping of molecules}
\begin{figure}[t!]
\centering
\includegraphics[width=\columnwidth]{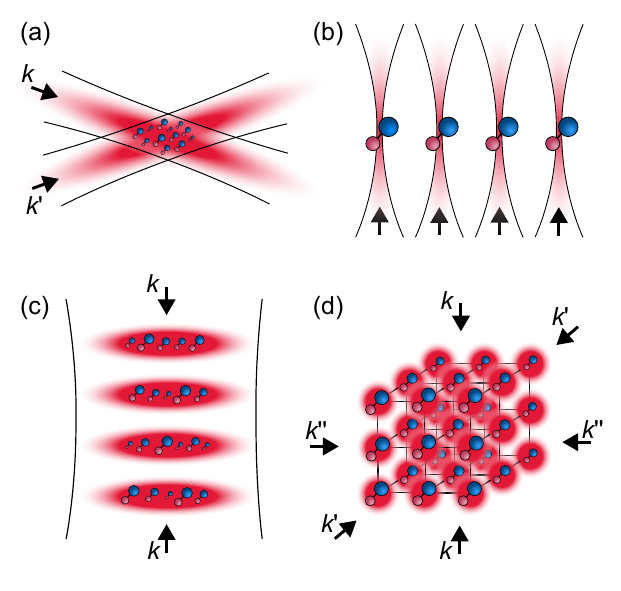}
\caption{Examples of different geometries of optical trap. In each diagram the arrow indicates the direction of the wavevector $k, k', k''$ of the light. (a)~3D optical dipole trap from the overlap of two beams. The two beams must be frequency shifted from each other to avoid interference effects. (b)~An array of tightly focused optical tweezers. (c)~A 1D optical lattice formed from the standing wave resulting from two same-frequency counter-propagating beams. (d)~A 3D optical lattice resulting from three pairs of counter-propagating beams.}
\label{fig:TrapGeometries}
\end{figure}

Optical traps are a workhorse of modern atomic physics~\cite{GrimmReview}, and play an integral part of the molecule-based quantum simulator. Optical traps can provide tight and controlled confinement for ultracold molecule gases in a range of geometries as illustrated in Fig.~\ref{fig:TrapGeometries}. Critically, they provide spatial confinement on length scales where dipolar interactions between neighbouring molecules are significant and take discrete values (for example, multiples of the lattice spacing in a 3D optical lattice).

In its simplest form, an optical trap works using a laser that is far detuned from an electronic transition such that the molecule feels a spatially dependent potential $U_\mathrm{trap}(\boldsymbol{r})$ that is proportional to the light intensity $I(\boldsymbol{r})$ such that
\begin{equation}
U_\mathrm{trap}(\boldsymbol{r}) = - \frac{1}{2\epsilon_0 c} \alpha I(\boldsymbol{r}).
\label{eq:DipoleTrap}
\end{equation}
The term of proportionality $\alpha$ is the dynamic polarisability of the molecule. This term is wavelength dependent as it is varies with the detuning of the light from the many electronic transitions that are available~\cite{Vexiau2017}. For linearly polarised trap light, the polarisability can be written as a sum over $i$ transitions
\begin{multline}
\alpha = \sum_{i}\frac{3\pi c^2\Gamma_i}{2 \omega_i^3\Delta_i} \times \\
\left( \frac{d^2_{\downarrow\uparrow(+)}}{2}\sin^2\beta + \frac{d^2_{\downarrow\uparrow(-)}}{2}\sin^2\beta + d^2_{\downarrow\uparrow(0)}\cos^2\beta  \right).
\end{multline}
Here $\omega_i$ is the resonant angular frequency of the transition $i$, and $\Delta_i$ is the detuning of the trap light from that transition. $\Gamma_i$ is the linewidth associated with the \emph{vibrational} transition dipole moment. The components $d_{\downarrow\uparrow(0)}, d_{\downarrow\uparrow(\pm)}$ correspond to the dipole moment on the \emph{rotational} $\pi, \sigma^{\pm}$ transitions respectively (given by Eq.~\ref{eq:dipMatrix}), and the angle $\beta$ is the angle of the trap light polarisation with respect to the quantisation ($z$) axis. 

When the light is sufficiently far detuned that the rotational structure of the transitions can be ignored, the polarisability becomes solely dependent on the $(N, M_N)$ of the state occupied by the molecules such that
\begin{equation}
\alpha_{N. M_N}(\beta) = \alpha^{(0)} + \alpha^{(2)} C_{N, M_N} P_2(\cos\beta)
\end{equation}
where
\begin{equation}
C_{N, M_N}=\frac{N(N+1)-3M_N^2}{(2N+3)(2N-1)}
\end{equation}
and $P_2(\cos\beta)$ corresponds to the associated Legendre polynomial $(3\cos^2\beta-1)/2$. The terms $\alpha^{(0)}$ and $\alpha^{(2)}$ quantify the scalar and tensor components of the polarisability and are often referred to as the isotropic and anisotropic polarisabilities, respectively. They are related to the molecule-frame polarisabilities in the directions parallel ($\alpha_\parallel$) and perpendicular ($\alpha_\perp$) to the bond axis such that
\begin{equation}
\alpha^{(0)} = \frac{1}{3}(\alpha_\parallel+2\alpha_\perp),
\end{equation}
\begin{equation}
\alpha^{(2)} = \frac{2}{3}(\alpha_\parallel-\alpha_\perp).
\label{eq:AnisotropicPolarisability}
\end{equation}
For molecules in the rotational ground state $N=0$, $C_{0,0}=0$ and therefore $\alpha_{0,0}=\alpha^{(0)}$. That is, the polarisability of the molecule becomes independent of the laser polarisation; this is due to the spherical symmetry of the wavefunction of the rotational ground state. For molecules in rotationally excited states however, the potential experienced by the molecules depends upon $N, M_N$ and the angle of the laser polarisation. 

To examine further, we define the ac Stark Hamiltonian that describes the interaction between the molecule and the oscillating electric field $\boldsymbol{E}_\mathrm{ac}(t)=\boldsymbol{E}_\mathrm{ac}\cos{\omega t}$ of the off-resonant trap light
\begin{equation}
\hat{H}_\mathrm{ac} = -\frac{1}{2} \boldsymbol{E}_\mathrm{ac}(t) \cdot \boldsymbol{\alpha} \cdot \boldsymbol{E}_\mathrm{ac}(t).
\end{equation}
For scalar polarisabilties, this expression is often presented as $-\frac{1}{4}\alpha |\boldsymbol{E}_\mathrm{ac}|^2$ where the extra factor of $\frac{1}{2}$ comes from time averaging of the electric field. The magnitude of $\boldsymbol{E}_\mathrm{ac}$ is related to the intensity of the light by $I=\frac{1}{2}\epsilon_0 c|\boldsymbol{E}_\mathrm{ac}|^2$, leading to the expression given in Eq.~\ref{eq:DipoleTrap}.

\begin{figure}[t!]
\centering
\includegraphics[width=\columnwidth]{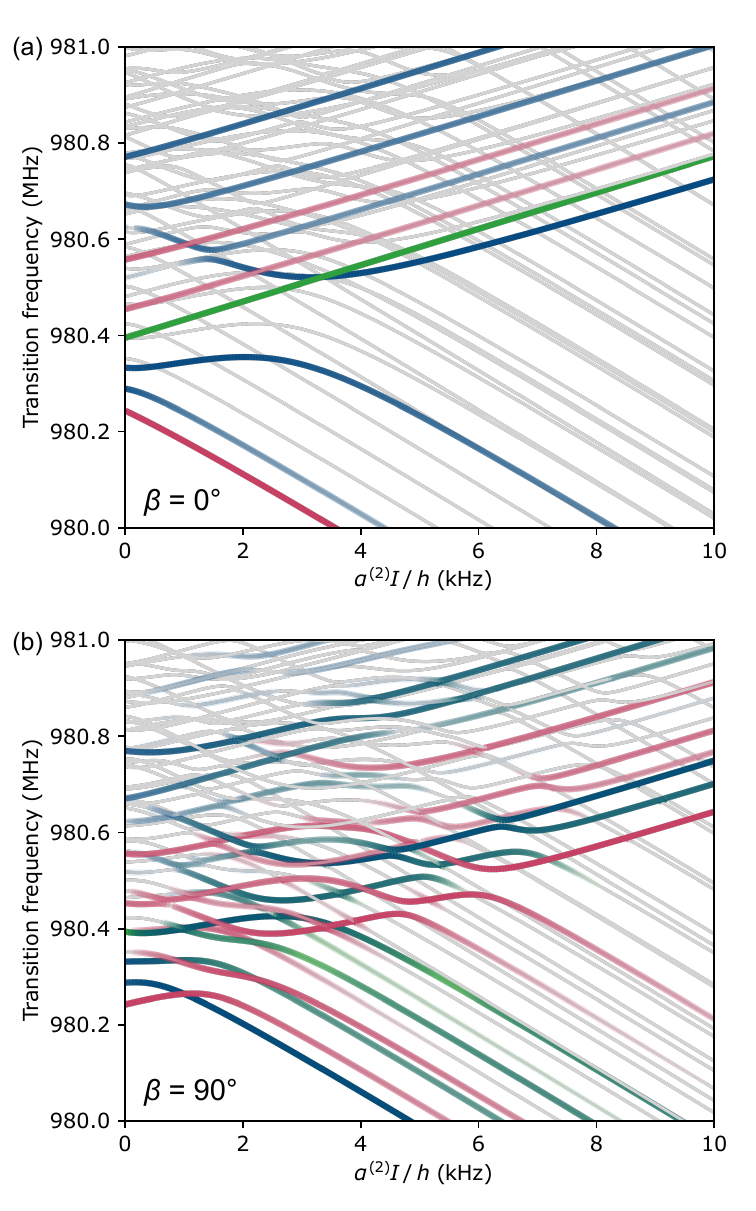}
\caption{Shift in the transition frequency from the ground hyperfine state $(0,5)_0$ to $N=1$ as a function of $\alpha^{(2)}I$. The light is linearly polarised at an angle (a)~$\beta=0^{\circ}$, (b)~$\beta=90^{\circ}$ to a 181.5\,G magnetic field applied along $z$. The states are colour coded in the same way as Fig.~\ref{fig:RbCs_electricfield}, such that red corresponds to transitions driven by microwaves polarised along $z$ and green/blue correspond to microwaves circularly polarised in the $xy$-plane such that they drive $\sigma^\pm$ transitions in the absence of the trap light.}
\label{fig:LightShift}
\end{figure}

In the $N, M_N$ basis $\hat{H}_\mathrm{ac}$ has elements
\begin{multline}
\bra{N,M_M}\hat{H}_\mathrm{ac}\ket{N',M'_N} = \alpha^{(0)}I\delta_{N,N'}\delta{M_N,M'_N} \\
+ \alpha^{(2)}I\sum_M D^2_{M0}(0,\beta,0) \sqrt{(2N+1)(2N'+1)} (-1)^{M_N} \\
\times \begin{pmatrix}
N & 2 & N'\\
-M_N & M & M'_N
\end{pmatrix}
\begin{pmatrix}
N & 2 & N'\\
0 & 0 & 0
\end{pmatrix}
\label{eq:acMatrix}
\end{multline}
where $D^2_{M0}$ is a Wigner D-matrix. Evaluating these matrix elements for $N=1$ yields
\begin{multline}
\bra{1, M_N} \hat{H}_\mathrm{ac} \ket{1, M'_N} = I \alpha^{(0)} + \frac{I\alpha^{(2)}}{5} \times \\
\begin{blockarray}{ccccc}
\bra{1,+1} & \bra{1,0} & \bra{1,-1} &\\
\begin{block}{(ccc)cc}
   -P_2(\cos\beta) & -\frac{3}{\sqrt{2}}\sin\beta\cos\beta & -\frac{3}{2}\sin^2\beta & & \ket{1,+1}\\
   -\frac{3}{\sqrt{2}}\sin\beta\cos\beta & 2P_2(\cos\beta) & \frac{3}{\sqrt{2}}\sin\beta\cos\beta & & \ket{1, 0}\\
   -\frac{3}{2}\sin^2\beta & \frac{3}{\sqrt{2}}\sin\beta\cos\beta & -P_2(\cos\beta) & .& \ket{1,-1}\\
\end{block}
\end{blockarray}
\label{eq:acMatrixN1}
\end{multline}
We can see from Eq.~\ref{eq:acMatrix} that all of the $N=1$ states are shifted by the isotropic polarisability, and this is indeed the case for all $N$; the isotropic component of the polarisability contributes to optical trapping but shifts all rotational states equally. However, the rotationally excited state also experiences linear shifts depending on $M_N$ and $\beta$ indicated by the associated Legendre polynomials on the main diagonal. The off-diagonal terms indicate couplings between different $M_N$ states. While such coupling may be negligible for molecules with large separation between hyperfine states, it is very important in molecules with small hyperfine energy splitting. For RbCs, the interaction between the molecular anisotropic polarisability and the trap light introduces significant additional mixing in the $N, M_N$ basis. We can observe these effects by examining the light shifts associated with the $N=0\rightarrow 1$ transition in RbCs~\cite{Gregory2017} as shown in Fig.~\ref{fig:LightShift}. 

When $\beta=0^\circ$ all of the off-diagonal terms in Eq.~\ref{eq:acMatrixN1} become zero. We therefore observe in Fig.~\ref{fig:LightShift}(a) a structure with predominantly linear shifts in energy with the occasional avoided crossing between states that have the same $M_F$ (and therefore components that have the same $M_N, m_\mathrm{Rb}, m_\mathrm{Cs}$). Because for this value of $\beta$ the non-zero matrix elements in Eq.\,\ref{eq:acMatrixN1} only couple states that have the same $M_F$, this remains a good quantum number. 

When $\beta=90^\circ$ on the other hand, there are non-zero off-diagonal terms that mix states with $\Delta M_N=\pm2$ in addition to the diagonal terms that cause linear shifts and couplings between states with $\Delta M_N=0$.The structure in this case is much more complex as shown in Fig.~\ref{fig:LightShift}(b). The additional mixing terms propagate through to the total angular momentum to couple states with $\Delta M_F=0,\pm2$ with the result being that $M_F$ is no longer a good quantum number. This can be viewed as the magnetic field and the linearly polarised trap light competing to be the effective quantisation axis; at large laser intensity the $M_F$ will be a good quantum number, but the relevant quantisation axis will no longer by $z$ but instead the direction that the laser polarisation is pointing. 

For all other values of $\beta$ all of the off-diagonal terms in Eq.~\ref{eq:acMatrixN1} are non-zero causing mixing between states with $\Delta M_F=0,\pm1,\pm2$. For RbCs this results generally in an even more complex structure. We will discuss later in the section on magic-polarisation traps the special case of $\beta=54.7^\circ$ where the diagonal terms of Eq.~\ref{eq:acMatrixN1} become zero leaving only the off-diagonal mixing terms.

\subsection*{Rotational coherence in optical traps}

Coherence is an important resource in quantum experiments as it defines the timescale over which quantum effects may be observed. A perceived loss of quantum coherence may be caused by any field that causes state-dependent shifts in energy that vary across the sample spatially or over time. This includes electric fields, magnetic fields, or the off-resonant optical fields associated with optical traps. For molecule experiments utilising the dipolar spin-exchange interactions, it is crucial that variations in the energy between states remain well below the dipolar interaction energy for the spin-exchange to remain resonant. When molecules are prepared in superpositions of rotational states, resulting in oscillating dipoles, the need to minimise variations in transition energy can also be viewed as necessary to keep the many oscillating dipoles all in phase with one another.  

\begin{figure}[t!]
\centering
\includegraphics[width=\columnwidth]{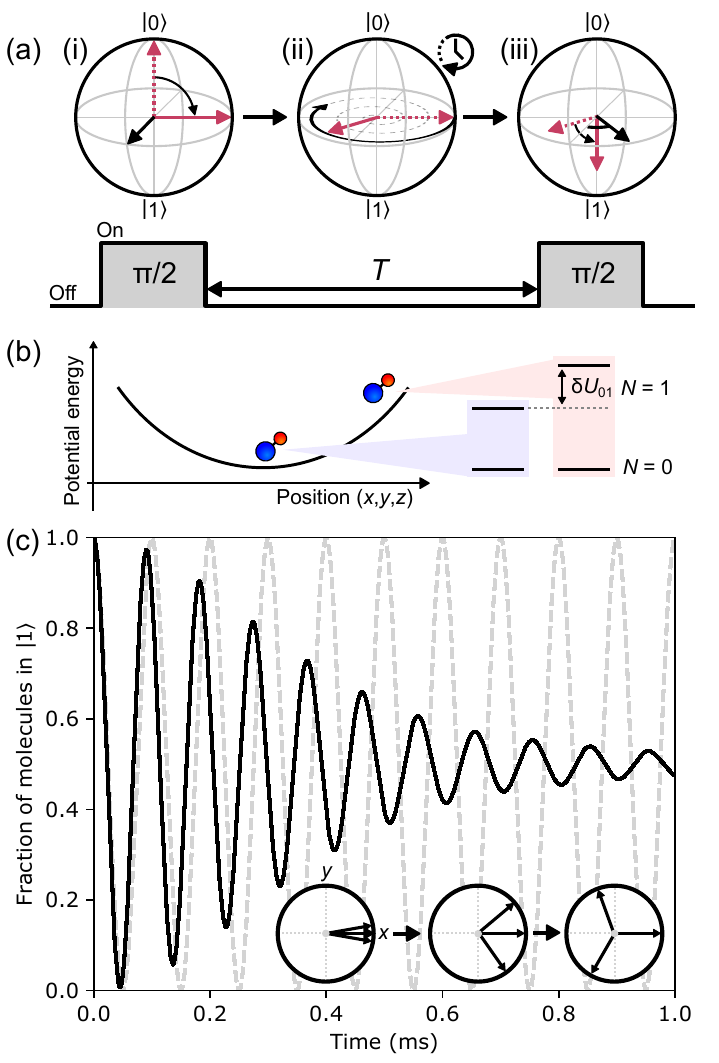}
\vspace{-0.7cm}
\caption{Dephasing in optical traps. (a)~Bloch sphere representation of Ramsey spectroscopy, along with the associated sequence of microwave pulses used in experiments. (b)~Diagram illustrating two molecules at different spatial positions in a common optical potential. The light experienced by each molecule depends on its position in the trap. In general, this causes a spatially dependent light shift of the transitions between rotational states. Here, the light shift for the $N=0\rightarrow 1$ transition is illustrated and labeled by $\delta E$. (c)~A simulated Ramsey measurement for 1000~molecules. The black solid line is the result for a 1\% difference in the polarisabilities of the two states. The grey dashed line shows the result for a `magic' trap that results in the same polarisability for each state. The calculation is performed for molecules that do not move during the measurement, but are thermally distributed following a Maxwell-Boltzmann distribution. The calculation assumes a harmonic potential with a trap frequency $\omega/2\pi=100$\,Hz, and a molecule sample that has a temperature of 1\,$\mu$K. This results in a spatial distribution that is Gaussian with a width $\sigma=9.8\,\mu$m. The molecules are displaced from the peak intensity of the light by gravitational sag $g/\omega^2=25\,\mu$m. To observe Ramsey fringes as a function of time, the microwaves are detuned from the transition by 10\,kHz. The inset shows an illustration of the dephasing in the Bloch sphere picture. We show three example vectors that start initialised in the same place, but spread out over the equatorial plane as they freely evolve at different rates.}
\label{fig:Decoherence}
\vspace{-0.3cm}
\end{figure}

The standard method to assess the lifetime of coherent superpositions is to use a Ramsey sequence which is illustrated with Bloch sphere diagrams in Fig.~\ref{fig:Decoherence}(a). For simplicity we consider specific labels for the transition between states $\ket{0}$ and $\ket{1}$. First, a $\pi/2$ pulse is performed to prepare the molecules in the coherent superposition $(\ket{0}+\ket{1})/\sqrt{2}$. The superposition is then allowed to evolve freely for a variable time $T$ during which time a phase $\mathrm{e}^{-iU_{01}t/\hbar}$ will be acquired, where $U_{01}$ is the energy difference between the superposed states. This free evolution is shown by the rotation of the Bloch vector in the horizontal plane. Finally, a second $\pi/2$ pulse is performed which projects the superposition with its acquired phase onto the $\ket{0},\ket{1}$ basis. We note that this general Ramsey sequence can be applied to any pair of states with the caveat that multiple photons may be required to drive transitions between states that are not directly connected by an allowed transition. 

Here, we will understand the problem with a worked example based on state-dependent light shifts in an optical trap that is particularly relevant to current experiments. Molecules in an optical trap formed from Gaussian beams experience variations in the intensity of the trap light as they move around the trap. If the two states in the superposition experience different intensity-dependent light shifts, the energy between the states will also vary depending on the position in the trap, as shown for two molecules in Fig.~\ref{fig:Decoherence}(b). The relative phase acquired by each molecule during the Ramsey free-evolution time is proportional to the energy difference the states, and so as the energy difference varies so does this phase acquired by the molecules. This results in the individual Bloch vectors associated with each molecule in the sample spreading out over the equator of the Bloch sphere; this indicates a loss of phase coherence across the sample, sometimes referred to as dephasing. As this happens, the Bloch vector averaged across the sample as a whole becomes smaller.

We observe this in experiments as a reduction in the contrast of Ramsey fringes as shown in Fig.~\ref{fig:Decoherence}(c). The phase acquired during the free evolution is measured with respect to that of the microwaves that perform the driving pulses. We therefore observe Ramsey fringes, oscillations in the number of molecules observed in each of the basis states, with a frequency equal to the detuning of the microwaves from resonance. The black solid line shows the expected Ramsey fringes for molecules experiencing an intensity-dependent light shift of the transition being probed; full details of the calculation are given in the caption. We see that the amplitude of the Ramsey fringes decays quickly, limiting the duration over which experiments can be usefully performed to $<1$\,ms. The grey dashed line shows shows the Ramsey fringes for molecules that do not experience any light shift on the transition. In this case the Ramsey fringes persist with full contrast. 

\subsection*{Dynamical decoupling}

\begin{figure}[t!]
\centering
\includegraphics[width=\columnwidth]{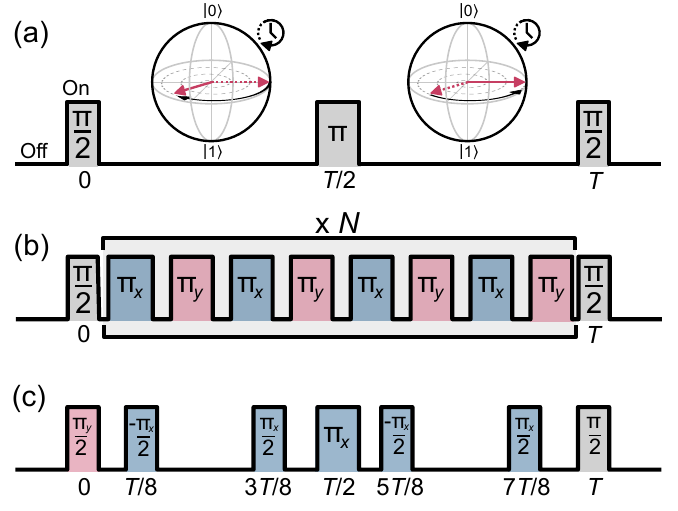}
\caption{Example pulse sequences used to implement dynamical decoupling. (a)~Spin echo. (b)~XY8. Here, a set of 8 equally-spaced $\pi$ pulses, with phase alternating by 90$^{\circ}$, is repeated $N$ times. (c)~WAHUHA. The times at which the pulses are performed are given below each of the pulses.}
\label{fig:CompositePulses}
\end{figure}

More complex sequences comprised of more pulses than the standard Ramsey sequence may be used to recover coherence or to maintain coherence for longer times. Such techniques were first developed for nuclear magnetic resonance (NMR) spectroscopy, but are now commonly applied for noise and error suppression in a wide range of quantum systems~\cite{Viola1999}. There are 3 example sequences that have been implemented in ultracold molecule experiments in Fig.~\ref{fig:CompositePulses}. 

Perhaps the simplest example of dynamical decoupling is the spin-echo sequence presented in Fig.~\ref{fig:CompositePulses}(a). Here the standard Ramsey sequence is modified to include a single $\pi$ pulse halfway through the free-evolution time. This pulse reverses the direction of precession of the Bloch vectors around the equator of the Bloch sphere. Provided that the average rate of precession for each vector is the same before and after the pulse, each Bloch vector should return to its initial state by the time of the final $\pi/2$ pulse. This sequence therefore cancels the effects of dephasing caused by different molecules precessing at different rates by canceling the precession of each molecule independently. In experiments using ultracold molecules, this approach has generally resulted in 1-2 orders of magnitude increase in the coherence time for spin-echo coherence compared to standard Ramsey sequences~\cite{Yan2013,Seesselberg2018,Burchesky2021}.

There are many variations of composite pulses that may be used in experiments. Periodic application of $\pi$ pulses, such as in the XY8 sequence shown in Fig.~\ref{fig:CompositePulses}(b) are more complex to implement but generally outperforms spin echo when the error in the pulses can be kept sufficiently small. Moreover, the Waugh-Huber-Haeberlen (WAHUHA) sequence shown in Fig.~\ref{fig:CompositePulses}(c) has particular utility as it enables suppression of pair-wise dipole-dipole interactions as well as sources of single-molecule dephasing~\cite{Yan2013}. Note that while composite pulses have many uses, it is important to keep the pulses short with respect to the overall free-evolution time, such that the system behaviour approximates the unperturbed system that is desired. In addition, it is often important for interactions that the single-particle dephasing be cancelled throughout the sequence; this is better approximated by the XY8 sequence than the more basic spin echo. In many situations it is best if experiments are able to eliminate light shifts altogether, or suppress them so that the dynamical decoupling pulses are a minority of the free-evolution of the system.

\subsection*{Magic-polarisation traps}

For optical traps, the light shift associated with a given pair of states may be suppressed by varying properties of the trap light, such as tuning the polarisation, intensity, or wavelength of the light. A trap that eliminates the light shift is commonly referred to as a `magic' trap. We will now give an overview of these techniques, starting with a magic trap that is realised by tuning the polarisation of the trap light. 

We have shown through Eq.~\ref{eq:acMatrixN1} that the tensor light shifts that lead to dephasing of rotational coherence are dependent on the laser polarisation. For linearly polarised light, the light shift can be tuned by varying the angle of the laser polarisation $\beta$. We have already touched on the laser polarisation of $\beta=54.7^\circ$ that will cancel all of the shifts that are linear with intensity (the diagonal elements in Eq.~\ref{eq:acMatrixN1}). However, as shown in Fig.~\ref{fig:LightShift}(c), quadratic shifts may still be present due to the combination of nuclear-spin hyperfine structure and mixing in the uncoupled basis (from the off-diagonal elements in Eq.~\ref{eq:acMatrixN1}). By tuning away from $\beta=54.7^\circ$, it is however possible to combine the linear and quadratic terms to give a turning point at the average intensity experienced by molecules in the trap. At this turning point, the intensity-dependent light shifts are not zero but are minimised.

\begin{figure*}[t!]
\centering
\includegraphics[width=2\columnwidth]{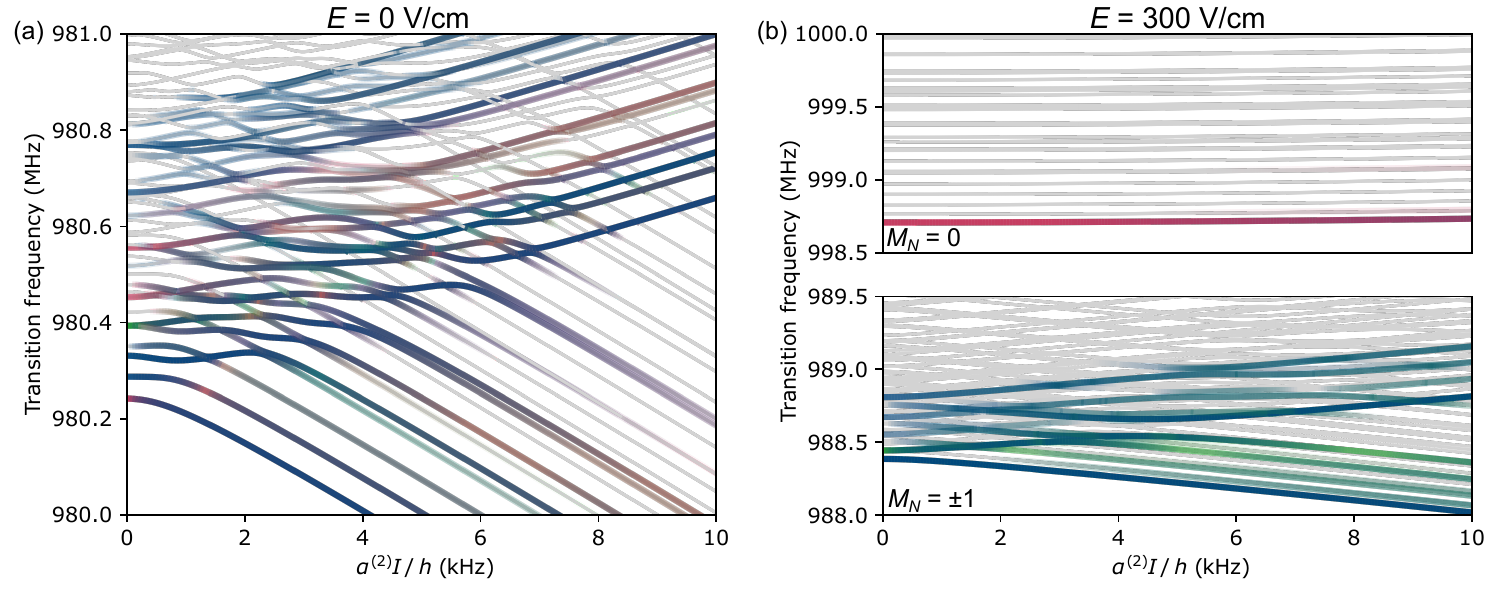}
\caption{Shift in the transition frequency from the ground hyperfine state $(0,5)_0$ of $^{87}$Rb$^{133}$Cs to $N=1$ as a function of $\alpha^{(2)}I$ for light that is linearly polarised at the magic angle $\beta=54.7^\circ$. (a)~With no electric field applied, couplings between states that differ in $M_N$ by $\Delta M_N=0, \pm1, \pm2$ cause a complex structure of avoided crossings and energy shifts. (b)~Light shift with the same polarisation trap light but with an electric field of 300\,V/cm. Now, the $M_N=0$ states separate from those with $M_N=\pm1$ as shown previously in Fig.\,\ref{fig:RbCs_electricfield}. As the $M_N=0$ states are now far from any levels that they may otherwise couple to, the effects of these couplings become diminished. }
\label{fig:MagicLightShift}
\end{figure*}

This approach was first demonstrated in KRb. Here, a coherence time of $\sim1.5$\,ms in Ramsey spectroscopy between $(N=0, M_N=0)\leftrightarrow(N=1,M_N=0)$ states for linearly polarised light with a magic angle of $\beta=48(4)^\circ$~\cite{Neyenhuis2012}. This represented an order of magnitude improvement in the coherence time compared to that achieved in the same trap with $\beta\sim0^\circ$ laser polarisation. Crucially the magic angle they observe depends on the light intensity, magnetic field, and electric field applied to the molecules. For superpositions of $M_N=0$ states, the magic polarisation may become more effective as larger electric fields are applied~\cite{Blackmore2020}. This is because the electric field separates the $M_N=0$ components from the $M_N$ components, reducing the coupling effects from the off-diagonal terms in the Hamiltonian. To illustrate this effect, a comparison of the light shift experienced by $^{87}$Rb$^{133}$Cs for light that is linearly polarised at the magic angle with and without a dc electric field is shown in Fig.~\ref{fig:MagicLightShift}. Similarly, the magic angle approach is highly effective in molecules such as CaF where large hyperfine and Zeeman splitting limits the effects of the couplings between hyperfine states~~\cite{Burchesky2021}. 

The magic angle is not practical in traps with very high laser intensity where the off-diagonal couplings dominate. In this case the linear polarisation of the trap light effectively becomes the relevant quantisation axis. The angle~$\beta$ therefore no longer affords the same degree of tunability. This can be seen in all panels of Fig.~\ref{fig:LightShift} where at the highest intensities, there are two bands of transitions, that either increase or decrease in energy at similar rates independent of the laser polarisation. 

An alternative technique for high-intensity optical trapping is to use elliptically polarised trap light~\cite{Rosenband2018}. In this case, the polarisation vector is chosen such that one component is $\pm i \sqrt{2}$ times that of the orthogonal component. In the limit of high intensity, this magic ellipticity causes one eigenstate of the light-shift-dominated Hamiltonian to become independent of the laser intensity. This has been tested with single NaCs molecules confined in optical tweezers. Coherence times of 62(3)\,ms were achieved with a spin-echo sequence, that increased to 250(40)\,ms with an XY8 sequence~\cite{Park2023}.

\subsection*{Magic-wavelength traps}

Varying the wavelength of the trap light changes the values of the polarisabilities $\alpha^{(0)}, \alpha^{(2)}$. This allows the possibility for engineering magic-wavelength traps, and has been explored in experiments with various molecules such as NaK~\cite{Bause2020} and RbCs~\cite{Gregory2024}. 

One strategy to implement a magic wavelength trap relies on the fact that the light shifts that depend on the rotational state that result from Eq.~\ref{eq:acMatrix} are \emph{all} proportional to the anisotropic component of the polarisability $\alpha^{(2)}$. If we can engineer an optical trap where $\alpha^{(2)}=0$ then it follows that we can eliminate both the linear and quadratic differential light shifts, at all intensities, simultaneously. From Eq.~\ref{eq:AnisotropicPolarisability}, we can see that $\alpha^{(2)}$ depends on the difference between the molecule-frame polarisabilities $\alpha_\parallel, \alpha_\perp$; when these two components are the same, the orientation of the molecule is no longer important and the dynamic polarisability becomes spatially isotropic.

The molecule-frame polarisabilities are related to a sum over the contributions from different electronic transitions in the molecule. The parallel component $\alpha_\parallel$ depends on the sum over all $\Sigma\rightarrow\Sigma'$ electronic transitions, while the perpendicular component $\alpha_\perp$ depends on the sum over $\Sigma\rightarrow\Pi'$ transitions~\cite{Vexiau2017}. In general, tuning the wavelength of the light allows us to change $\alpha_\parallel, \alpha_\perp$ and therefore $\alpha^{(0)}, \alpha^{(2)}$ as this coarsely changes the detuning of the light from the various electronic transitions; a magic-wavelength trap may therefore be found that sets $\alpha^{(2)}=0$. 

\begin{figure}[t!]
\centering
\includegraphics[width=1.05\columnwidth]{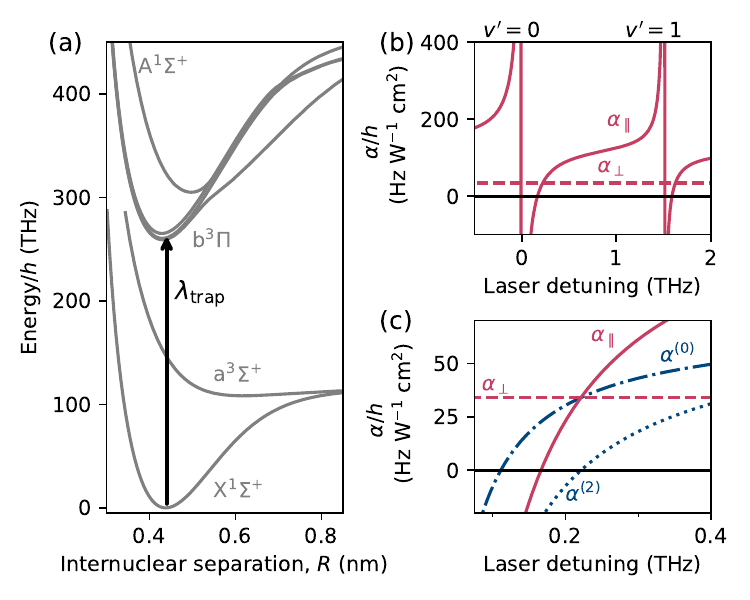}
\caption{Magic wavelength trapping of RbCs using the lowest $\mathrm{X}^1\Sigma\rightarrow\mathrm{b}^3\Pi$ transitions. (a)~Energy potentials associated with the RbCs molecule. The black arrow indicates the energy of the magic-wavelength trap light that is tuned close to transitions to the lowest energy levels of the $\mathrm{b^3\Pi_0}$ state. (b)~Parallel and perpendicular components of the polarisability ($\alpha_\parallel, \alpha_\perp$) as a function of laser detuning in the region around the transitions to the two lowest energy vibrational states $v'=0,1$. (c)~Zoom in to the region around $\sim200$\,GHz blue-detuned from $v'=0$ where $\alpha_\parallel \approx \alpha_\perp$. The resulting isotropic and anisotropic components ($\alpha^{(0)},\alpha^{(2)}$) are also shown.}
\label{fig:MagicWavelength}
\end{figure}

The theoretical basis for this type of magic-wavelength trapping of ground-state bialkali molecules was described by Guan $et~al.$~\cite{Guan2021}. Here, the light is tuned close to a nominally-forbidden transition from $\mathrm{X}^1\Sigma$ to the lowest energy states of $\mathrm{b}^3\Pi$; this transition appears forbidden as it changes the electronic angular momentum $S=0\rightarrow1$. However, the transition is weakly-allowed due to mixing between the $\mathrm{b}^3\Pi$ and nearby $\mathrm{A}^1\Sigma$ potentials shown in Fig.~\ref{fig:MagicWavelength}(a) for RbCs, such that even the lowest-energy states in $\mathrm{b}^3\Pi$ have some small amount of $\mathrm{A}^1\Sigma$ character. 

Tuning the light close to these transitions effectively accesses a weak $^1\Sigma\rightarrow^1\Sigma$ transition that only tunes $\alpha_\parallel$. The variation in $\alpha_\parallel$ around the transitions to the two lowest vibrational state $v'=0,1$ of $\mathrm{b}^3\Pi$ are shown in Fig.~\ref{fig:MagicWavelength}(b). At the location of each transition $\alpha_\parallel$ strongly diverges to $\pm\infty$ depending on whether the light approaches the transition as red or blue detuned. For light tuned between the transitions to these two vibrational states, the value of $\alpha_\parallel$ can be chosen to be any value. The value of $\alpha_\perp$ on the other hand is effectively a constant value over this range of laser frequencies, as the light is far-red detuned from the nearest $^1\Sigma\rightarrow^1\Pi$ transition by $\sim100$\,THz. 

In Fig.~\ref{fig:MagicWavelength}(c) we can see that $\alpha_\parallel\approx\alpha_\perp$ when the trap light is blue detuned by $\sim200$\,GHz from the transition to $v'=0$. When these two values are matched $\alpha^{(2)}$ becomes zero, while there is still a non-zero and positive value of $\alpha^{(0)}$ that enables optical trapping at regions of high laser intensity. Note that a laser detuning of 200\,GHz is effectively far detuned from these weak, and therefore narrow linewidth, transitions; the linewidth of the nearest transition is $\Gamma_{v'=0}=14.1(7)$\,kHz~\cite{Das2026}. This results in in detuning on the order of 10$^{8}\Gamma$ such that trap loss due to photon scattering is low enough  trap loss due to photon scattering to allow long trap lifetimes.

This scheme has been demonstrated with RbCs molecules in bulk gases and optical tweezer arrays. For bulk gases, Ramsey coherence times of 0.78(4)\,s and spin-echo coherence times in excess of 1.4\,s were achieved in the absence of dipole-dipole interactions~\cite{Gregory2024}. For tweezer-confined molecules coherence times of 15(6)\,s were measured~\cite{Ruttley2024}, and the scheme extended to observe second-scale coherences between three levels simultaneously~\cite{Hepworth2025}. 

While the scheme at first glance seems generally applicable to any bialkali molecule, it is best suited to the heaviest of the bialkalis. This is because the scheme relies on the assumption that the trap light is sufficiently far detuned that the rotational structure of the $\mathrm{b}^3\Pi$ states can be neglected. When this is not the case, $\alpha_\parallel$ and therefore $\alpha^{(0)},\alpha^{(2)}$ become dependent on $N$~\cite{Hepworth2025}. Heavier molecules experience stronger coupling between $\mathrm{b}^3\Pi$ and $\mathrm{A}^1\Sigma$ states, resulting in stronger transitions which enables larger detuning. They also possess smaller rotational constants leading to smaller separation in energy between the rotational states. Even for RbCs, the heaviest of the bialkali molecules, the presence of rotational structure in the excited state leads to small deviations from the ideal case described. This results in $\sim100$\,MHz variation in the detuning required to make the trap rotationally magic depending on $N, M_N, N', M'_N$ and the laser polarisation~\cite{Gregory2024,Ruttley2024,Hepworth2025}. 

Prior to the work by Guan~$et~al.$~\cite{Guan2021}, magic-wavelength trapping had been demonstrated in $^{23}$Na$^{40}$K~\cite{Bause2020} using the same $\mathrm{X}^1\Sigma\rightarrow\mathrm{b}^3\Pi$ transition. However, as the molecule is much lighter, the approach was different with the trap not setting the $\alpha^{(2)}\approx 0$, but instead tuning the values of $\alpha^{(0)}, \alpha^{(2)}$ such that two chosen states of the molecule experience the same polarisability for a fixed laser polarisation and electric field. A similar scheme has also been identified for NaRb based on an experimental spectroscopic study of the lowest $\mathrm{b}^3\Pi$ states~\cite{He2021}.

\section{Ultracold molecular collisions}

An understanding of collisions between ultracold molecules is necessary for the study of bulk samples, or in weak optical lattices where molecules are allowed to tunnel between sites. Collisions are also important for evaporative cooling of molecules, which is necessary for the preparation of colder and denser samples. 

Ultracold polar molecules are typically categorized as either reactive or nonreactive depending on whether there is an energetically-allowed atom-exchange reaction between a pair of molecules. For a generic diatomic molecule $\mathrm{XY}$, comprised of two different nuclei $\mathrm{X}$ and $\mathrm{Y}$, there are three possible atom-exchange reactions that could be undergone. Two of these reactions lead to the formation of trimers 
\begin{equation*}
\mathrm{XY} + \mathrm{XY} \rightarrow \mathrm{X_2Y}+\mathrm{Y},
\end{equation*}
\begin{equation*}
\mathrm{XY} + \mathrm{XY} \rightarrow \mathrm{X}+\mathrm{XY_2}.
\end{equation*}
These reactions are always energetically forbidden, that is, the energy of the trimer and atom combined is greater than that of the two diatomic molecules. The remaining atom-exchange reaction 
\begin{equation}
\mathrm{XY} + \mathrm{XY} \rightarrow \mathrm{X_2}+\mathrm{Y_2}.
\end{equation}
converts the pair of heteronuclear molecules into two homonuclear molecules. This reaction is energetically-allowed for reactive molecules, and not allowed for nonreactive molecules~\cite{Zuchowski2010}; perhaps the most intensively studied reactive molecule is KRb, while commonly studied nonreactive molecules include NaRb, NaK, and RbCs. Most experiments only detect the specific species of molecules that are prepared, and then often only one particular hyperfine level of that molecule, and so the conversion of molecules to a different species will be observed as loss from the trap. The naive expectation is that molecules that are nonreactive should experience a lower rate of collisional loss than reactive molecules as there is no obvious mechanism for two-body loss from the trap. 

Surprisingly, experiments have found that exceedingly high loss rates associated with two-body collisions are observed regardless of the energetics of these reactions~\cite{Ye2018}. The loss rates observed for all molecules studied so far are at or close to the two-body universal rate constant. This rate constant is the limit in which all pairs of molecules are lost with unit probability whenever the two molecules are sufficiently close together to experience the short range part of their interaction potential, defined by the van der Waals length~\cite{Idziaszek2010}. This limit can therefore be seen as the case where all pairs of molecules are lost from the sample when they experience a collision. For nonreactive molecules, understanding the mechanism that leads to molecule-pair loss with near-absolute certainty is an ongoing challenge for the field. 

One possible mechanism for fast loss of nonreactive molecules was originally proposed by Mayle~\emph{et~al}~\cite{Mayle2012,Mayle2013}. They suggested that when the two molecules collide they access a dense structure of Feshbach resonances. This corresponds to a large number of rovibrational states, bound by the molecular potentials associated with the 4-atom molecule, that have the same energy as the incoming molecule pair. Scattering in this highly-resonant regime leads to the formation of long-lived collision complexes; in effect the molecules become stuck together for a time while they explore the rotational and vibrational states of the 4-atom molecule. This has led to the mechanisms colloquial name of `sticky collisions'. A simple conceptual analogy put forward by John Bohn compares the situation to an insect flying into a bottle through the narrow neck. The insect buzzes around inside the bottle bumping into the walks randomly until it eventually finds the neck again. The `sticking time' is then the period it takes the insect to escape. The narrower the neck and the bigger the bottle, the longer this takes.

The lifetime of a collision complex $\tau_\mathrm{c}$ is related to the width $\Gamma$ of the resonant state by $\tau_\mathrm{c}=\hbar/\Gamma$. For a dense structure of resonances, the mean resonance width can be estimated using Rice-Ramsperger-Kassel-Marcus (RRKM) theory~\cite{Levine2005} to find the expected complex lifetime in the highly-resonant regime
\begin{equation}
\tau_\mathrm{c}=\frac{h\rho}{N_0}
\end{equation}
where $\rho$ is the density of the coupled states and $N_0$ is the number of energetically-open channels for the free-molecule pair. For nonreactive molecules $N_0=1$, and for reactive molecules $N_0\gg1$. The expected lifetime associated with an $(^{23}\mathrm{Na}^{39}\mathrm{K})_2$ complex has been calculated to be 6\,$\upmu$s, and the lifetimes for other bialkali collision complexes found to range from 0.25\,$\upmu$s to 253\,$\upmu$s by extrapolation using approximate scaling laws~\cite{Christianen2020dos}. 

\begin{figure}[t!]
\centering
\includegraphics[width=\columnwidth]{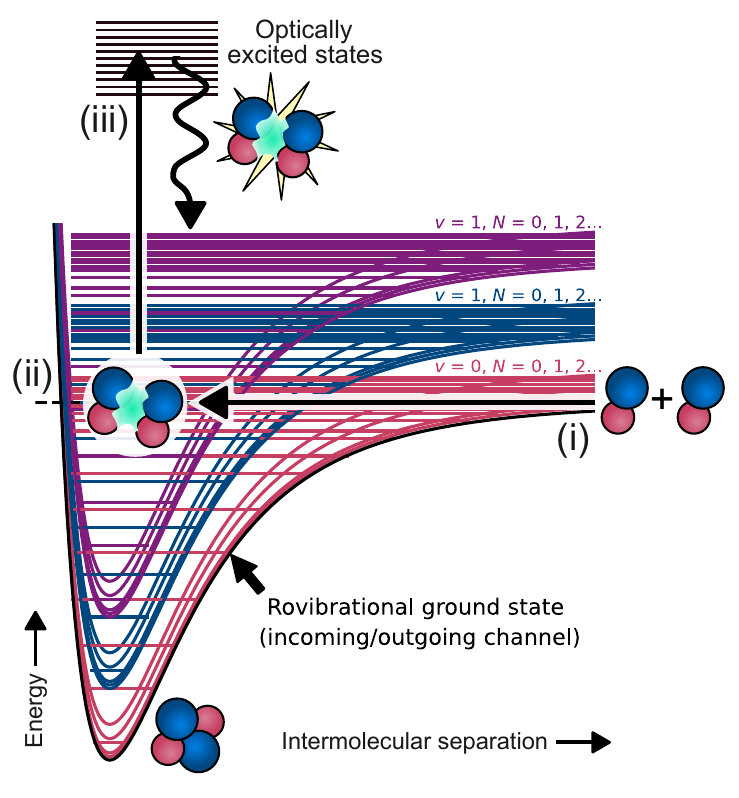}
\caption{Loss of molecules by optical excitation of long-lived two-molecule collision complexes in bright optical traps. (i)~A pair of molecules collide. During this collision there is a chance that the molecules will become close enough together to experience the short-range part of their intermolecular potential. (ii)~The molecules become stuck together for a time due to the dense structure of rovibrational bound states that are accessible. (iii)~While the molecules are stuck together, they may be optically excited by the trap light. They will then decay from these excited states to a different state to which they originally started and be observed as lost.}
\label{fig:StickyCollisions}
\end{figure}

The formation of collision complexes on its own does not result in permanent loss from the trap; the molecules may not be detectable while in the complex, but they will become unstuck after a time and become available for detection again. However permanent loss of molecules can occur during a sticky collision when there is an additional mechanism that causes fast loss of molecules in the collision complex. Christianen~$et~al$ proposed that electronic excitation by the high laser intensities present in optical traps may cause loss in this way~\cite{Christianen2020photo}. They determined theoretically that the expected time for photon scattering by the $(^{23}\mathrm{Na}^{39}\mathrm{K})_2$ complex is $\sim1\,\upmu$s across a br~oad range of trap laser wavelengths. This is much shorter than the lifetime of the complex in the dark, and therefore molecules entering collision complexes are unlikely to exit the complex before scattering a photon. This photon-scattering process will change the state of the molecules, such that they occupy an electronically excited state of the complex that can then decay to any number of states in the 4-atom molecule. The combined mechanism of optical excitation of long-lived two-body collision complexes is depicted in Fig.~\ref{fig:StickyCollisions} and can explain the observations of fast two-body losses in nonreactive molecule samples.

Pioneering experiments performed with reactive KRb molecules by the Ni group use ionization spectroscopy and velocity-map imaging in order to detect the reactants, intermediates, and products of reactive collisions in ultracold samples of KRb directly. They have directly observed the formation of collision complex intermediates~\cite{Hu2019}, and also shown that even for reactive molecules, photoinduced loss of complexes can be the dominant source of collisional loss. By placing the molecules in the dark, they were able to directly measure the lifetime of the complex to be 0.36(3)\,$\upmu$s in agreement with RRKM theory~\cite{Liu2020}. They also studied nonreactive two-body collisions of molecules with atoms, where the complex lifetime should be much shorter. However, in this case they measured a complex lifetime that was $10^5$ times longer than that predicted by RRKM~\cite{Nichols2022}.

Experiments to verify that this mechanism takes place in nonreactive molecule-molecule collisions have been much more limited in their detection capabilities as they only detect the reactants in the specific hyperfine state that they are initially prepared. The general approach has been to place the molecules in the dark for some or all of the trapping time and to search for suppression of the collisional loss associated with the dark time. So far, the RbCs molecule is the only nonreactive molecule where this process has been able to be verified. Experiments were performed by placing the molecules in an optical trap where the intensity is periodically modulated with time. A larger number of molecules were observed when the molecules were placed in the dark for 75\% of the total trap time. By varying the duration of the dark periods and fitting the variation in molecule number to a rate equation model, the lifetime of the $(\mathrm{RbCs})_2$ complex in the dark was found to be 0.53(6)\,ms, when the molecules were prepared in their ground hyperfine state, within a factor of 2 of that predicted using RRKM theory~\cite{Christianen2020dos}. Though a later study showed that the observed suppression of loss varied with the hyperfine state that the molecules were prepared in, indicating either some unexpected variation in the lifetime of the complex or in the photoexcitation rate~\cite{Gregory2021collisions}. 

A broader understanding of collisional loss of nonreactive molecule-molecule collisions still remains to be found. Many experiments have been performed across other nonreactive bialkali molecules, with attempts to measure the lifetimes of the associated collision complexes either by also using intensity-modulated traps~\cite{Gersema2021}, or for example using blue-detuned box traps~\cite{Bause2021}. These experiments find significant disagreement with RRKM predictions, suggesting complex lifetimes that are many orders of magnitude greater than expected. As such, there is currently no complete explanation for the loss of nonreactive molecules and as such this is an open area of research.

\subsection*{Collisions in lattices}

Suppression of collisional loss can be achieved by preparing molecules in optical lattices provided that there is only one molecule per lattice site. In this case, a tightly-confining lattice that suppresses tunneling will result in no collisions as molecules remain on their respective sites. For lattices that are weak enough to allow tunneling, there is the possibility of two molecules entering the same site and therefore experience fast loss via the two-body collision. However, this mechanism for loss may be suppressed via the continuous quantum Zeno effect, as reported for KRb molecules~\cite{Zhu2014}. This suppression of loss in weak optical lattices arises precisely due to the fast two-body loss, which corresponds energetically to a strong on-site interaction. There is therefore a significant energy penalty associated with two molecules entering the same site of the optical lattice, which dramatically reduces the likelihood of this occurring. When discussing tunneling in lattices, the case where particles are unable to enter the same site is often described as the hardcore boson limit. Ultracold molecules by default are well described by kinetics in this limit.

\subsection*{Collisional shielding}
\label{sec:shielding}

\begin{figure}[t!]
\centering
\includegraphics[width=1.05\columnwidth]{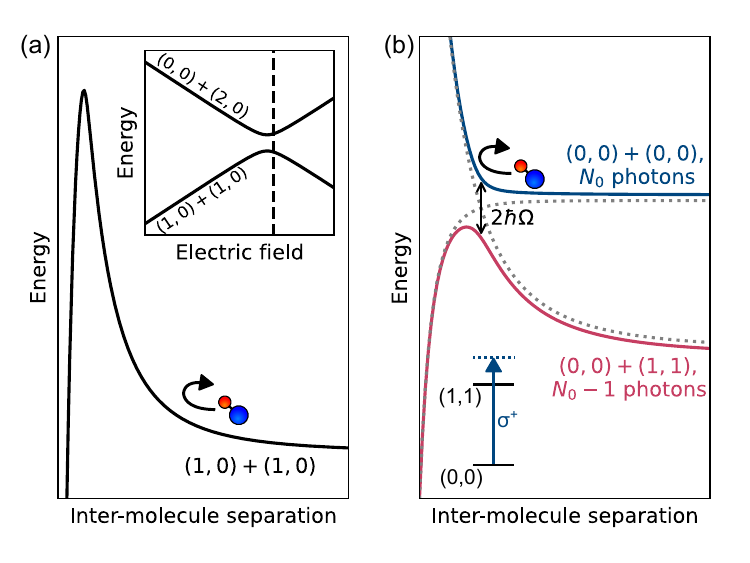}
\vspace{-0.8cm}
\caption{Schematic of the collisional shielding approach with (a)~dc electric fields and (b)~circularly-polarised microwaves. The inset of (a) shows the dc Stark shifts of the pair states $(N=0, M_N=0)+(2,0)$ and $(1,0)+(1,0)$, where the states become degenerate in energy there is an avoided crossing. With the electric field just above the avoided crossing, indicated by the vertical dashed line, the molecules experience a repulsive barrier shown in the main panel. In (b)~molecules are prepared in the ground state $(0,0)$ and a circularly polarised microwave field, blue-detuned from the transition to $(1,1)$ is applied as shown inset to prepare the molecules in a microwave-dressed state that also experiences a repulsive barrier.  }
\label{fig:Shielding}
\end{figure}
An alternative way to suppress collisional loss is to engineer a long-range repulsive barrier between the molecules using dipole-dipole interactions. Such a barrier stops pairs of molecules from entering the short-range part of the potential where they will otherwise become stuck, effectively shielding them from the two-body loss. By enabling elastic collisions between molecules (i.e. collisions that do not result in loss) samples are able to efficiently thermalise and be evaporatively cooled. For this reason, collisional shielding of polar molecules has been instrumental in the recent realisations of quantum-degenerate gases of polar molecules~\cite{DeMarco2019,Schindewolf2022, Bigagli2024,Shi2025}. 

The first collisional shielding was observed by applying dc electric fields to a sample of fermionic KRb molecules in a 1D optical lattice~\cite{deMiranda2011}. The 1D lattice restricts the motion of the molecules to a 2D plane, and then the electric field oriented the dipoles orthogonal to that plane such that molecules preferentially approach each other via repulsive side-by-side collisions. This approach was later extended to 3D using a resonant effect shown in Fig.~\ref{fig:Shielding}(a); the molecules are prepared in the single state $(N=1, M_N=0)$ and the electric field is tuned such that the energy of two $(1,0)$ molecules becomes resonant with the energy of a pair of molecules in the different states $(0,0)+(2,0)$. Coupling between these two pair-states results in an avoided crossing as a function of electric field, centred at an electric field of 3.2$B/d$. When the electric field is set to be just above the avoided crossing the pairs of colliding molecules experience an effective repulsion regardless of their relative trajectories as they enter the collision~\cite{Matsuda2020,Li2021}. 

Collisional shielding may also be implemented using off-resonant microwaves. This approach was first observed for collisions between pairs of CaF molecules~\cite{Anderegg2021}, but has now also been implemented across various species of bialkali molecules. In its simplest form, microwave shielding works with molecules prepared in $N=0$, and uses circularly-polarised microwaves to address the $\sigma^+$ transition to $(N=1,M_N=1)$. When the microwaves are blue detuned from resonance, pairs of molecules colliding in $N=0$ couple to a repulsive state corresponding to a molecule pair with one molecule in $N=0$ and the other in $N=1$, which forms the collisional shield~\cite{Karman2018, Lassabliere2018} as shown in Fig.~\ref{fig:Shielding}(b).

For molecules shielded with only a circularly polarised microwave field, the intermolecular potential has an attractive component at long-range that can support bound states. The energies of these bound states can be tuned by varying the microwave field and give rise to field-linked resonances that can dramatically tune the collisional properties of the molecules~\cite{Avdeenkov2003,Avdeenkov2004,Chen2023}. For bosonic molecules, it was found that the presence of these bound states causes 3-body loss at high densities, preventing cooling to quantum degeneracy. However, it was subsequently shown that these 3-body losses can be negated by adding a second microwave driving field, polarised to drive the $\pi$ transitions that couple $N=0$ to the $(N=1, M_N=0)$ states. This second microwave field allows the attractive component to be cancelled to realise a potential that is only repulsive~\cite{Karman2025}. This development was critical to the realisation of molecular Bose-Einstein condensates~\cite{Bigagli2024,Shi2025}.

Molecules undergoing double-microwave shielding experience experience a time-averaged interaction that consists of two components;  a repulsive van der Waals interaction potential that scales as $1/r^6$, and a longer-range dipole-dipole interaction that scales as $1/r^3$~\cite{Zhang2026, Karman2025}. The $\pi$ and $\sigma^+$ driving fields result in dipoles oscillating in orthogonal directions, either vertically or in the horizontal plane respectively, and therefore their time-averaged interactions have opposite signs. The dipole-dipole interactions may therefore be tuned between attractive and repulsive, or turned off altogether. Moreover, this summation allows the tuning of the form of the spatial anisotropy of the interactions. While the shielding must be maintained to continually suppress loss of the molecules, there is a broad range of parameter space where the shielding is sufficiently effective while allowing exquisite control over the strength and anisotropy of their interactions. Developing these techniques further and their application in experiments is now an exciting area of research.

\section{Quantum spin dynamics with polar molecules}
\label{sec:QuantumSimulation}

Ultracold polar molecules have a promising future as a platform for the quantum simulation of many-body quantum spin dynamics~\cite{Cornish2024}. We illustrate their potential by showing how interactions in tweezers and lattices can lead to rich physics and quantum entanglement. The general approach follows that of the Ramsey sequence given in the context of rotational dephasing in section\,III. However, we will now introduce dipole-dipole interactions to the sequence; when the interactions are sufficiently strong, the molecules become entangled during the evolution time. To understand this, we will start with perhaps the simplest situation of two molecules prepared side-by-side in optical tweezers.

\subsection*{Entangling pairs of molecules}

We consider a pair of molecules with each molecule in one of two rotational states that we map onto a spin-1/2 system of pseudospins $\ket{\downarrow}\equiv\ket{N=0, M_N=0},\ket{\uparrow}\equiv\ket{1,1}$. For a single molecule the $\pi/2$ pulse operation is described by the rotation matrix
\begin{equation}
R_x(\pi/2) \equiv \frac{1}{\sqrt{2}}
\begin{pmatrix}
1 &  -i \\
-i &  1
\end{pmatrix}.
\end{equation}
The molecule pair on the other hand is described in the pair state basis 
\begin{equation}
\{\ket{\uparrow},\ket{\downarrow} \}\otimes\{\ket{\uparrow},\ket{\downarrow}\} = \{\downarrow\downarrow,\uparrow\downarrow,\downarrow\uparrow,\uparrow\uparrow\}
\end{equation}
for which the two-molecule rotation matrix is
\begin{equation}
M_{\pi/2} \equiv R_x(\pi/2)\otimes R_x(\pi/2) = \frac{1}{2} 
\begin{pmatrix}
1 & -i & -i & -1 \\
-i & 1 & -1 & -i \\
-i & -1 & 1 & -i \\
-1 & -i & -i & 1 \\
\end{pmatrix}.
\end{equation}
During the Ramsey free-evolution time, the dynamics of the molecule-pair is governed by the spin-exchange Hamiltonian given in Eq.\,\ref{eq:SpinHamiltonian} which for our choice of basis states reduces to 
\begin{multline}
H_\mathrm{dd}^{q=0} = \frac{1-3\cos^2\Theta}{4\pi\epsilon_0r^3} \left(\frac{J_\perp}{2}\left(\hat{S}^{(i)}_{+}\hat{S}^{(j)}_- + \hat{S}^{(i)}_-\hat{S}^{(j)}_+\right)\right) \\
\equiv 
\frac{1-3\cos^2\Theta}{4\pi\epsilon_0r^3}\left(\frac{J_\perp}{2}\right)
\begin{pmatrix}
0 & 0 & 0 & 0 \\
0 & 0 & 1 & 0 \\
0 & 1 & 0 & 0 \\
0 & 0 & 0 & 0 \\
\end{pmatrix}.
\end{multline}
During the evolution time $T$ the pair-state therefore evolves as
\begin{equation}
M_\mathrm{int} \equiv \mathrm{e}^{-i H_\mathrm{dd}^{q=0} T / \hbar} = 
\begin{pmatrix}
1 & 0 & 0 & 0 \\
0 & \cos(4\pi\tilde{T}) & -i\sin(4\pi\tilde{T}) & 0 \\
0 & -i\sin(4\pi\tilde{T}) & \cos(4\pi\tilde{T}) & 0 \\
0 & 0 & 0 & 1 \\
\end{pmatrix}
\end{equation}
where $\tilde{T} = \frac{1-3\cos^2\Theta}{4\pi\epsilon_0 r^3} \left( \frac{J_\perp T}{4}\right)$. The evolution associated with the complete Ramsey sequence is given by the product of the evolution matrices
\begin{equation}
M \equiv M_{\pi/2} \cdot M_\mathrm{int} \cdot M_{\pi/2},
\end{equation}
and for molecules initialised first in $\ket{\downarrow\downarrow}$, the final state is given by
\begin{multline}
\ket{\Phi(T)} = M\ket{\downarrow\downarrow} \\
= -\mathrm{e}^{-2\pi i \tilde{T}}\left[ \cos(2\pi\tilde{T})\ket{\uparrow\uparrow} - i \sin(2\pi\tilde{T})\ket{\downarrow\downarrow} \right].
\label{eq:EntangledState}
\end{multline}
At times $\tilde{T}=1/8, 3/8, 5/8...$ the molecules become maximally entangled as Eq.\,\ref{eq:EntangledState} becomes an equal superposition of $\ket{\downarrow\downarrow}$ and $\ket{\uparrow\uparrow}$, equivalent to a Bell state with an imaginary phase. This demonstrates how a Ramsey pulse sequence can generate entanglement between pairs of molecules. Suddenly stopping the interaction at exactly this time will leave the molecules in this entangled state, and controlled entanglement using this scheme has been demonstrated recently in a range of experiments with tweezer-confined molecules~\cite{Holland2023, Bao2023, Picard2025, Ruttley2025}.   

\subsection*{Ensembles of polar molecules in optical lattices}

Larger ensembles of molecules may be entangled using the same Ramsey procedure, though the evolution becomes non-trivial to calculate analytically. Also crucially for molecules in optical lattices detection of both the state and position in the lattice becomes important. Recent advances have seen techniques borrowed from \emph{atomic} quantum gas microscopy applied to ultracold molecules enabling the precise read-out of both the state and position of molecules in the array~\cite{Covey2018, Rosenberg2022, Christakis2023, Mortlock2026}. 

Spin-1/2 molecules in optical lattices are naturally described in the hardcore boson limit by the \tjvw Hamiltonian~\cite{Gorshkov2011prl, Gorshkov2011pra}
\begin{equation}
\begin{split}
& \hat{H}_{tJVW} = - t\sum_{\left<i,j\right>,m}{(c^{\dagger}_{im}c_{jm}+c_{im}c^{\dagger}_{jm})}\\ 
&+ \sum_{i\neq j}{\frac{(k/2)}{|{\bf R}_{i}-{\bf R}_j| ^{3}}} \times\bigg[ \frac{J_\perp}{2} (S^+_i S^-_j+S^-_i S^+_j) + J_z S^z_i S^z_j \\ 
& ~~~~~~~~~~~~~~~~~~~~~~~~~~~~~~~~~~~{+ Vn_i n_j + W (n_i S^z_j + n_j S^z_i) \bigg]}.
\end{split}
\label{eq:tjvw}
\end{equation}
The first term describes tunneling of molecules between lattice sites at a rate $t/\hbar$. The creation and annihilation operators $c^\dagger_{j,m}$, $c_{j,m}$ are defined such that they add or remove a molecule on site $j$ in state $m$. The second term describes dipole-dipole interactions between molecules on different sites, with the strength scaled by the distance $|{\bf R}_i-{\bf R}_j|$ between sites $i$~and~$j$. The term $k=1/(4\pi\epsilon_0)$ is Coulomb's constant, and this is divided by 2 to compensate for the double counting of pairs in the sum over~$i,j$. 

The dipole-dipole interaction term can be further broken down by the various contributions given in the brackets. The $J_\perp$ and $J_z$ terms describe the spin-exchange interactions and spin-spin interactions between the space-fixed dipoles. These terms have been described in isolation previously and are given in Eq.~\ref{eq:SpinHamiltonian}. The final two components describe spin-density and density-density interactions associated with the dipoles, where $n_i$ describes the occupation of the site~$i$. The density-density interaction coefficient $V=(d_\downarrow+d_\uparrow)^2/4$, that is it depends on the sum of the lab-frame dipoles of the molecules on the two sites. The spin-density interaction coefficient $W=(d_\downarrow^2-d_\uparrow^2)/2$ on the other hand depends on the difference in the magnitude of the two lab-frame dipoles. 

The magnitude of all of the terms in Eq.~\ref{eq:tjvw} can be precisely tuned in experiments. The rate of tunneling $t/\hbar$ can be controlled independently of the others as it alone varies with the depth of the optical lattice; control of the lattice depth is achieved experimentally by varying the intensity of the trap light. For very large trap depth, the tunneling rate will become vanishingly small and the dynamics of Eq.\,\ref{eq:tjvw} become equivalent to an Ising spin model. The various dipole-dipole interactions on the other hand are somewhat more coupled to each other. Varying the applied electric field modifies all of these terms at the same time. The response of the terms relative to each other then depends on the combination of states chosen for the spin mapping. It is the relative values that are most important for experiments as it is the interplay between terms and the shifting of dominance from one term to another that can give rise to new quantum phases and phase transitions. 

It is possible to selectively remove selected dipole-dipole interaction terms by lowering their strength such that they are negligible; if all of the interaction terms are removed from Eq.\,\ref{eq:tjvw} reduces to the hardcore limit of the Hubbard model. The $J_\perp$ term can be removed by mapping onto a pair of states with no dipole-allowed $\sigma^\pm$ transition between them, similarly $J_z$ terms can be removed by ensuring no dipole-allowed $\pi$ transitions or electric fields are applied. Both $V$ and $W$ terms could be removed by operating in a lattice with unity filling such that there are no variations in the density; this is however difficult to achieve in experiments with the current maximum lattice filling fractions being around $50\%$. Instead, the use of dynamical decoupling removes the effect of $W$~\cite{Carroll2025}. By combining dynamical decoupling with operation at relatively low field such that $V$ is small, experiments have so far focused on examining the region where $J_\perp$ and $J_z$ alone are the dominant sources of dipole-dipole interactions~\cite{Li2023, Carroll2025}. 

Another approach to engineering the interactions between molecules is to use a Floquet approach~\cite{Eckardt2017}. Here, transitions in the molecule are periodically driven using microwave pulses, with the period of the modulation being much shorter than the typical interaction timescale in the system. The time-averaged interactions may therefore be different from that of the original Hamiltonian. This was recently used, for example, to engineer both $J_z$ and $J_\perp$ in NaRb without requiring any dc electric fields~\cite{Christakis2023}. 

The \tjvw Hamiltonian is expected to exhibit many discrete quantum phases. Gorshkov\,$et~al.$ performed numerical calculations for a 1D chain of molecules as a function of the filling fraction ($n$) and the ratio $J_\perp/t$ with $V=W=J_z=0$, and found evidence for a rich quantum phase diagram exhibiting spin-density waves, superfluid behaviours, and phase separation~\cite{Gorshkov2011prl}. This indicates exciting prospects for future experiments with lattice-confined polar molecules that are just reaching the point where controlled realisations of this Hamiltonian are possible. 

To conclude, there are many rich possibilities for the study of many-body dynamics with polar molecules. The rich phase diagram associated with arrays of molecules is now within experimental reach, underpinned by important recent developments in loading~\cite{Reichsollner2017, Holland2023, Bao2023, Picard2024, Ruttley2024, Walraven2024,Baldock2026,Karman2026loading}, site-resolved detection of molecular arrays~\cite{Rosenberg2022,Christakis2023,Mortlock2026}, and in the precise engineering of dipole-dipole interactions via Floquet engineering~\cite{Schuster2021, Christakis2023, Miller2024}. Recent reports of quantum-degenerate gases of polar molecules~\cite{DeMarco2019,Schindewolf2022,Bigagli2024,Shi2025} along with the precise control over collisions via collisional shielding represent a paradigm shift for the field. Such quantum-degenerate samples are anticipated to possess a rich phase diagram including ordered structured phases even without the need for optical lattices~\cite{Schmidt2022}. We hope that this article provides the basic understanding needed to explore these topics in greater depth. 

This work is the author's original manuscript that has since been published in Contemporary Physics (2026) and is available at \href{https://doi.org/10.1080/00107514.2026.2738291}{DOI:10.1080/00107514.2026.2738291}.

\section{Acknowledgements}
We acknowledge support from the UK Engineering and Physical Sciences Research Council (EPSRC) Grants EP/P01058X/1, EP/W00299X/1, EP/W016141/1, EP/Y01510X/1 and UKRI2226 funded through the Programme Grant Scheme, UK Research and Innovation (UKRI) Frontier Research Grant EP/X023354/1, the Royal Society, and Durham University. PDG is supported by a Royal Society University Research Fellowship (URF/R1/231274).
\bibliography{ref}

\end{document}